\documentclass[preprint,12pt]{elsarticle}

\usepackage{amssymb}
\usepackage{amsmath}
\usepackage{amsthm}
\usepackage{dsfont}
\usepackage{subfig}
\usepackage{algorithm}
\usepackage{algpseudocode}
\newmuskip\pFqmuskip

\newcommand*\pFq[6][8]{%
  \begingroup 
  \pFqmuskip=#1mu\relax
  \mathchardef\normalcomma=\mathcode`,
  \mathcode`\,=\string"8000
  \begingroup\lccode`\~=`\,
  \lowercase{\endgroup\let~}\pFqcomma
  {}_{#2}F_{#3}{\left[\genfrac..{0pt}{}{#4}{#5};#6\right]}%
  \endgroup
}
\newcommand{\pFqcomma}{{\normalcomma}\mskip\pFqmuskip}

\newtheorem*{theorem*}{Theorem}

\usepackage{comment}

\usepackage{color}

\journal{Journal of Computational Physics}

\begin{document}

\begin{frontmatter}



\title{An adaptive lattice Green's function method for external flows with two unbounded and one homogeneous directions. }


\author[inst1]{Wei Hou\corref{cor1}}
\ead{whou@caltech.edu}
\author[inst1]{Tim Colonius}
\ead{colonius@caltech.edu}
\cortext[cor1]{Corresponding author}

\affiliation[inst1]{organization={Dept. Mechanical and Civil Engineering, Caltech},
            addressline={1200 E California Blvd}, 
            city={Pasadena},
            postcode={91125}, 
            state={California},
            country={USA}}

\begin{abstract}
We solve the incompressible Navier-Stokes equations using a lattice Green's function (LGF) approach, including immersed boundaries (IB) and adaptive mesh refinement (AMR), for external flows with one homogeneous direction (e.g. infinite cylinders of arbitrary cross-section).  We hybridize a Fourier collocation (pseudo-spectral) method for the homogeneous direction with a specially designed, staggered-grid finite-volume scheme on an AMR grid.  The Fourier series is also truncated variably according to the refinement level in the other directions.  We derive new algorithms to tabulate the LGF of the screened Poisson operator and viscous integrating factor. After adapting other algorithmic details from the fully inhomogeneous case \cite{yu2021multi}, we validate and demonstrate the new method with transitional and turbulent flows over a circular cylinder at $Re=300$ and $Re=12,000$, respectively.

\end{abstract}




\end{frontmatter}


\section{Introduction}

The lattice Green's function (LGF) is the analytical inverse of a discrete elliptic operator on an unbounded domain (lattice). 
Due to its value in numerical applications \citep{liska2016fast, cserti2000application}, its computation \citep{buneman1971analytic, martinsson2002asymptotic, katsura1971lattice} and asymptotic behavior \citep{martinsson2002asymptotic, katsura1973asymptotic} have been studied thoroughly.  In particular, for incompressible flows, the LGF can be combined with the immersed boundary method (IB) \cite{peskin2002immersed} to create an efficient and parallel algorithm \cite{liska2016fast}.  Efficiency can be further enhanced with a multi-resolution LGF framework for adaptive mesh refinement (AMR) \cite{dorschner2020fast,yu2021multi}.  In these methods, LGF is used to solve the pressure-Poisson equation and/or vorticity-streamfunction equation. The use of LGF ensures that the solution is defined everywhere in the free space (without imposition of artificial boundary conditions) yet only a finite set of active cells is needed to time-step the flow.  This yields a snug, adaptive domain that encloses only the evolving vorticity field, truncated at a small threshold value at the boundary.

However, LGF methods have only been developed for fully unbounded three-dimensional flow domains. In practice, a wide class of interesting geometries and flows exhibit span-wise periodicity. Among these flows are the flow past bodies with infinite spans and a constant two-dimensional cross-section, such as circular cylinders and unswept airfoils. 
Although a wide variety of other methods have been developed for these flows (e.g. \cite{dong2005dns, mittal1997inclusion, lehmkuhl2013low}), the multi-resolution LGF method promises greater computational efficiency while exactly preserving the asymptotic structure of the irrotational outer solution.

In the current study, we extend the multi-resolution framework that combines LGF, IB, and AMR \cite{yu2021multi} to solve fully 3D flows with one homogeneous direction. We exploit the spanwise periodicity by using a Fourier expansion of the flow variables (velocity, pressure, and IB forcing) and derive the evolution equations of the corresponding Fourier coefficients. This formulation enables us to compute the nonlinear convective term efficiently via the (dealiased) Fast Fourier Transform (FFT).  We develop a staggered-grid strategy that hybridizes the second-order finite-volume discretization for the inhomogenous directions with the Fourier expansion in the homogeneous one, while maintaining desired discrete conservation and other mimetic properties associated with the original 3D finite-volume discretization. In addition, we adaptively truncate Fourier coefficients to make the spanwise resolution consistent with the finite-volume AMR grid. 

With one periodic direction, the pressure is determined by a discrete screened Poisson equation for which we derive formulas and algorithms to efficiently evaluate the LGF. This particular LGF poses a unique challenge in two aspects. First, the discrete screened Poisson equation involves a continuous coefficient such that the corresponding LGF varies nonlinearly with it. A large number of these LGFs would thus be required, and it is desirable to have a fast way to evaluate them at runtime. While fast algorithms for certain LGFs exist \cite{buneman1971analytic,guttmann2010lattice,liska2014parallel,gabbard2024lattice}, the screened Poisson equation is not among them.  Moreover, in contrast to the regular Poisson equation, the existing polyharmonic asymptotic expansion does not apply to the LGF of the screened Poisson equation \cite{duffin1953discrete, duffin1958difference,martinsson2002asymptotic, gabbard2024lattice}. Thus, we need to directly compute the LGF through numerical integration. To address both challenges, we derive a numerically advantageous integral expression of this particular LGF and propose a spectrally convergent trapezoidal rule approximation. Similar challenges are also present in the handling of the viscous term. Thus, we derive and compute the LGF for the appropriate integrating factor for the viscous term. In addition, we provide algorithms for LGF of the integrating factor for the viscous Laplacian (the heat equation kernel), which allows us to employ a half-explicit Runge-Kutta (IF-HERK) method for time advancement \cite{liska2017fast}.


The paper is arranged as follows. We introduce the Fourier-transformed Navier-Stokes equations with IB forcing in Section~\ref{sec:FourierExp}. Then, in Section~\ref{sec:sp_disc}, we derive a spatial discretization in terms of corresponding discrete operators. In Section~\ref{sec:lgf}, we develop LGFs for the screened Poisson operator and integrating factor. In the next sections, we adapt several previous algorithms to the spanwise homogenous case, specifically the time marching method (Section~\ref{sec:time_disc}), multi-resolution application of the LGF (Section~\ref{sec:apply_lgf_mr}), and the domain and mesh adaptation strategies (Section~\ref{sec:adaptivity}).  The overall algorithm for the incompressible Navier-Stokes equations is then summarized in Section~\ref{sec:alg_sum}. Subsequently, in Section~\ref{sec:parallel}, we describe the parallelization strategy computational efficiency. In Section~\ref{sec:convergence}, we empirically demonstrate the (overall first-order) convergence of the scheme. Finally, in Section~\ref{sec:validation}, we validate the algorithm by computing flow past a cylinder at $Re=300$, and we highlight the ability of our algorithm to tackle large problems by computing the turbulent flow past a cylinder at $Re=12,000$.

\section{Governing equations and Fourier expansion}

\label{sec:FourierExp}

Physically, the problem under consideration is an infinite-span cylinder (axis $z$) of arbitrary cross-section moving (including acceleration) in the $x-y$ plane through an otherwise quiescent, viscous, incompressible fluid.  Invoking homogeneity, we restrict $z$ to a periodic section of a specified length $c$. For real-valued $f=f(x,y,z,t)$, we write the truncated Fourier series
\begin{align}
    f(x,y,z,t) & \approx \tilde{\boldsymbol{f}}_0 (x,y,t) + \sum_{k = 1}^{N/2} \left[\tilde{\boldsymbol{f}}_k (x,y,t) e^{2\pi i k z/c} + \overline{\tilde{\boldsymbol{f}}_k} (x,y,t) e^{-2\pi i k z/c}\right],
\end{align}
where 
\begin{align}
    \tilde{f}_k (x,y,t) = \mathcal{F}_k [f] := \frac{1}{c}\int^{c/2}_{-c/2} f(z)e^{-iz\frac{2\pi k}{c}} dz.
\end{align}

Let $\boldsymbol{u}$ and $\boldsymbol{\omega} = \nabla \times \boldsymbol{u}$ be the velocity and vorticity, and $p$ be the pressure, all nondimensionalized with respect to a specified velocity scale, length scale, and density.  In physical space, the incompressible Navier-Stokes equations with the IB formulation are \cite{liska2017fast}

\begin{equation}
    \begin{aligned} 
    \frac{\partial \boldsymbol{u}}{\partial t} + \boldsymbol{\omega} \times \boldsymbol{u}_a &= -\nabla \Pi + \frac{1}{Re} \nabla ^2 \boldsymbol{u} + \int_{\Gamma(t)} \boldsymbol{f}_\Gamma(\boldsymbol{\xi}, t) \delta(\boldsymbol{X}(\boldsymbol{\xi}, t) - \boldsymbol{x}) d\boldsymbol{\xi}, \\
    \nabla \cdot \boldsymbol{u} &= 0, \\
    \boldsymbol{u}_\Gamma(\boldsymbol{\xi},t) &= \int_{\mathbb{R}^3}\boldsymbol{u}(\boldsymbol{x},t)\delta(\boldsymbol{x} - \boldsymbol{X}(\boldsymbol{\xi}, t)) d\boldsymbol{x}.
    \end{aligned}
    \label{eq:NS_ib}
\end{equation}
Here, $\boldsymbol{x}$ and $\boldsymbol{x}_a = \boldsymbol{x} - \boldsymbol{R}(t)$ denote the coordinates in the inertial reference frame and those in the non-inertial reference frame, respectively. The non-inertial frame translates with the trajectory $\boldsymbol{R}(t)$ and rotates with angular velocity $\boldsymbol{\Omega}(t)$. $\boldsymbol{u}$ is the velocity vectors in the inertia reference frame. $\boldsymbol{u}_a$ is the velocity vector with respect to the non-inertial reference frame. The two velocities are related through $\boldsymbol{u} = \boldsymbol{u}_a + \boldsymbol{u}_r$ where $\boldsymbol{u}_r = \frac{d\boldsymbol{R}(t)}{dt} + \boldsymbol{\Omega(t)} \times \boldsymbol{x}_a := \boldsymbol{U}(t) + \boldsymbol{\Omega(t)} \times \boldsymbol{x}_a$. In this equation, $\frac{\partial}{\partial t}$ is the derivative in $t$ with $\boldsymbol{x}_a$ held constant, and $\nabla$ the gradients with respect to $\boldsymbol{x}_a$. Correspondingly, $\Pi = p- \frac{1}{2}|\boldsymbol{u}_r|^2 - \frac{1}{2}|\boldsymbol{u}_a|^2$ where $p$ is the pressure.

If we denote the immersed boundary points in the non-inertial frame as $\boldsymbol{X}_a(\boldsymbol{\xi}, t)$, we can rewrite the boundary condition
\begin{equation}
    \boldsymbol{u}_{\Gamma,a}(\boldsymbol{\xi},t) + \boldsymbol{U}(t) + \boldsymbol{\Omega}(t) \times \boldsymbol{X}_a(\boldsymbol{\xi}, t) = \int_{\mathbb{R}^3}\boldsymbol{u}(\boldsymbol{x},t)\delta(\boldsymbol{x} - \boldsymbol{X}(\boldsymbol{\xi}, t)) d\boldsymbol{x}.
\end{equation}
Note that the convolution integral is taken in the inertial coordinates. The Dirac delta function uses the relative position between the immersed boundary surface and the coordinates in the inertial reference frame. Thus, we are only sampling inertial frame velocity on the immersed boundary and equate the values to the inertial frame velocity boundary condition, on each single point parameterized by $\boldsymbol{\xi}$ and $t$.

In Fourier space, these equations read
\begin{equation}
\begin{aligned} 
\frac{\partial {\tilde{\boldsymbol{u}}}_k}{\partial t} + \mathcal{F}_k[\boldsymbol{\omega} \times \boldsymbol{u}_a] &= -\widetilde{\nabla}_k \tilde{\Pi}_k + \frac{1}{Re} \widetilde{\nabla ^2}_k \tilde{\boldsymbol{u}}_k\\
&+ \int_{\Gamma(t)_{2D}} \tilde{\boldsymbol{f}}_{\Gamma, k} (\boldsymbol{\xi}_{2D}, t)\delta_{2D}(\boldsymbol{X}_{2D}(\boldsymbol{\xi}_{2D}, t) - \boldsymbol{x}_{2D}) d\boldsymbol{\xi}_{2D},\\
    \widetilde{\nabla}_k . \tilde{\boldsymbol{u}}_k &= 0, \\
    \widetilde{\boldsymbol{u}}_{\Gamma, k}(\boldsymbol{\xi}_{2D}, t) &= \int_{\mathbb{R}^2}{\tilde{\boldsymbol{u}}}_k(\boldsymbol{x}_{2D},t)\delta_{2D}(\boldsymbol{x}_{2D} - \boldsymbol{X}_{2D}(\boldsymbol{\xi}_{2D}, t))d\boldsymbol{x}_{2D},
\end{aligned}
\label{eq:FT_NS_ib}
\end{equation}
where 
\begin{equation}
    \widetilde{\nabla}_k . \tilde{{\boldsymbol{u}}}_k = \frac{\partial \tilde{{{u}}}_k}{\partial x} + \frac{\partial \tilde{{v}}_k}{\partial y} + \frac{2\pi i k}{c}\tilde{w}_k,
\end{equation}
\begin{equation}
    \widetilde{\nabla}_k \tilde{\Pi}_k = \left[ \frac{\partial \tilde{\Pi}_k}{\partial x}, \frac{\partial \tilde{\Pi}_k}{\partial y}, \frac{2\pi i k}{c}\tilde{\Pi}_k \right]^T,
\end{equation}
\begin{equation}
    \widetilde{\nabla}_k^2  \tilde{{u}}_k = \frac{\partial^2 \tilde{u}_k}{\partial x^2} + \frac{\partial^2 \tilde{u}_k}{\partial y^2} - \left(\frac{2\pi k}{c}\right)^2\tilde{u}_k.
\end{equation}
Details for the Fourier transform of the IB terms are given in \ref{ap:NS_ib_Fourier}.

Since the body and flow are homogeneous in the $z$ direction, we simplify the boundary condition equations to
\begin{equation}
\begin{aligned} 
    \boldsymbol{u}_\Gamma(\boldsymbol{\xi}_{2D}, t) &= \int_{\mathbb{R}^2}{\tilde{\boldsymbol{u}}}_0(\boldsymbol{x}_{2D},t)\delta_{2D}(\boldsymbol{x}_{2D} - \boldsymbol{X}_{2D}(\boldsymbol{\xi}_{2D}, t))d\boldsymbol{x}_{2D} \quad k = 0,\\
    0 &= \int_{\mathbb{R}^2}{\tilde{\boldsymbol{u}}}_k(\boldsymbol{x}_{2D},t)\delta_{2D}(\boldsymbol{x}_{2D} - \boldsymbol{X}_{2D}(\boldsymbol{\xi}_{2D}, t))d\boldsymbol{x}_{2D} \quad k \neq 0.
\end{aligned}
\end{equation}

We can evaluate the nonlinear term efficiently using a de-aliased pseudo-spectral approach \cite{orszag1971numerical}, i.e. we discretize the inverse transform to the DFT, form the product in physical space via padded inverse transforms, and transform the product back to its truncated Fourier coefficients.  Padding via the $3/2$ rule is sufficient since the equations involve at most quadratic nonlinearity.

\section{Spatial Discretization}
\label{sec:sp_disc}

\begin{figure}
    \centering
    \includegraphics[width = 0.4\textwidth]{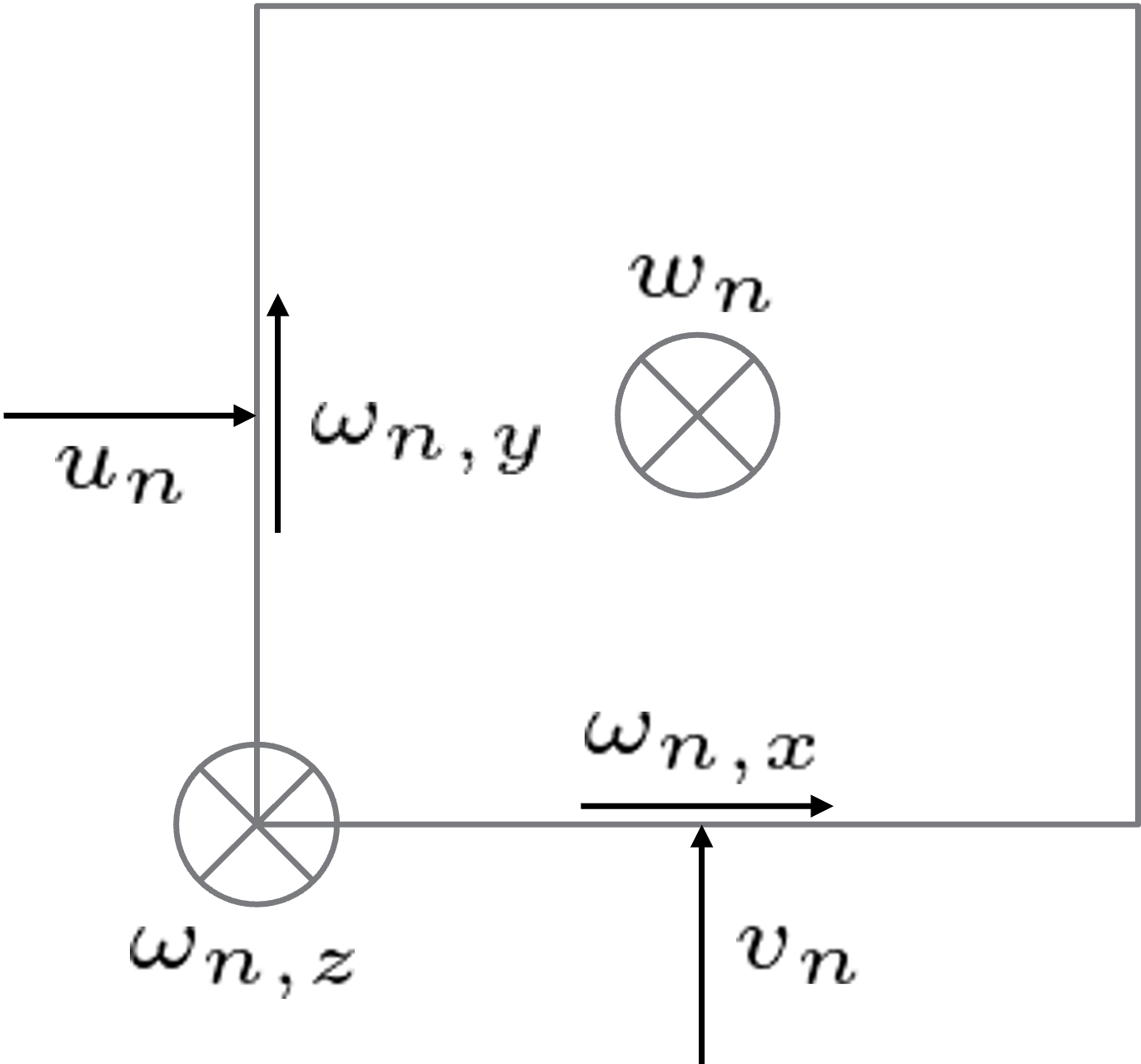}
    \caption{Variable placement in the $x-y$ plane for a Fourier interpolation sampling point in the $z$ direction.}
    \label{fig:VariablePlacement}
\end{figure}

To develop a framework that is best suited for the hybridized discretization, we place all finite-volume cells with their centers aligned with the evenly-spaced sampling points for Fourier interpolation.  The $x$ and $y$ components of velocity are placed on the faces, and the $z$ component of velocity is at the cell center.  One can visualize the resulting data field as a collection of evenly-spaced slices that appear as a 2D finite-volume method depicted in Figure~\ref{fig:VariablePlacement}.

With this staggered mesh, we now define discrete operators and enumerate some of their properties. In this section, we use boldfaced variables, e.g. $\boldsymbol{g} = [g_1, g_2, g_3]^T$, to denote 3-component vector fields and non-boldfaced variables, e.g. $g$, to denote scalar fields.  The operators are:
\begin{itemize}
    \item Divergence on $k^{th}$ Fourier coefficient:
    \begin{equation}
        D_k \boldsymbol{g} = \mathcal{D}_x g_1 + \mathcal{D}_y g_2 + (2\pi i k/c)g_3.
    \end{equation}
    \item Gradient on $k^{th}$ Fourier coefficient:
    \begin{equation}
        G_k g = [-\mathcal{D}_x^T g, -\mathcal{D}_y^T g, (2\pi i k/c)g]^T.
    \end{equation}
    \item Curl on $k^{th}$ Fourier coefficient: 
    \begin{equation}
        C_k \boldsymbol{g} = [-\mathcal{D}^T_y g_3 - (2\pi i k/c) g_2, (2\pi i k/c)g_1 + \mathcal{D}^T_xg_3, -\mathcal{D}^T_x g_2 + \mathcal{D}^T_y g_1]^T.
    \end{equation}
    \item Laplacian on $k^{th}$ Fourier coefficient:
    \begin{equation}
        L_k g = -\mathcal{D}_x^T\mathcal{D}_x g -\mathcal{D}_y^T\mathcal{D}_y g - (2\pi k /c)^2 g.
    \end{equation}
\end{itemize}
In the equations above, $\mathcal{D}$ denotes a forward finite difference derivative, for example, $\mathcal{D}_x g(i,j) = [g(i+1, j) - g(i, j)]/\Delta x$. The operators mimic some properties of the continuous operators, namely $D_k = -G^*_k$ and $L_k = -G^*_kG_k$ (where the superscript$\,^*$ denotes the conjugate transpose).

Let $\boldsymbol{u}_{a,n} = [u_n, v_n, w_n]^T$ be the velocity in the non-inertial frame at the $n^{th}$ slice and $\boldsymbol{\omega}_n = [\omega_{n,x}, \omega_{n,y}, \omega_{n,z}]^T$ be the vorticity on that slice, we discretize the nonlinear advection terms in rotational form by defining $\boldsymbol{N}_n(i,j) = [N_{n, x}(i,j),N_{n, y}(i,j),N_{n, z}(i,j)]^T$, and writing
\begin{equation}
\begin{aligned}
    N_{n,x}(i,j) = &\frac{1}{2}\omega_{n,y}(i,j)[w_n(i,j) + w_n(i-1,j)] \\
    - &\frac{1}{4} \omega_{n,z}(i,j)[v_n(i,j) + v_n(i-1,j)] \\
    - &\frac{1}{4}\omega_{n,z}(i,j+1)[v_n(i,j+1) + v_n(i-1,j+1)] ,\\
    N_{n,y}(i,j) = &\frac{1}{4} \omega_{n,z}(i,j)[u_n(i,j) + u_n(i,j-1)] \\
    + &\frac{1}{4}\omega_{n,z}(i+1,j)[u_n(i+1,j) + u_n(i+1,j-1)] \\
    - &\frac{1}{2}\omega_{n,x}(i,j)[w_n(i,j) + w_n(i,j-1)], \\
    N_{n,z}(i,j) = &\frac{1}{2}[\omega_{n,x}(i,j)v_n(i,j) + \omega_{n,x}(i,j+1) v_n(i,j+1)] \\
    - &\frac{1}{2}[\omega_{n,y}(i,j)u_n(i,j) + \omega_{n,y}(i+1,j) u_n(i+1,j)].
    \label{eq:dis_nonlin}
\end{aligned}
\end{equation}

Let $\boldsymbol{N}(\boldsymbol{\omega}, \boldsymbol{u})$ be the collection of the nonlinear advection term across all slices, evaluated using $\boldsymbol{\omega}, \boldsymbol{u}$ with Eq.~\ref{eq:dis_nonlin}. Inserting the discrete spatial operators in Eq.~\ref{eq:FT_NS_ib}, we obtain a system of (index 2) differential-algebraic equations (DAE)
\begin{equation}
    \begin{aligned} 
\frac{d {\tilde{\boldsymbol{u}}}_k}{d t} + \mathcal{F}_k[\boldsymbol{N}(\boldsymbol{\omega}, \boldsymbol{u})] &= -G_k \tilde{q}_k + \frac{1}{Re} L_k \tilde{\boldsymbol{u}}_k + P(t)^T \tilde{\boldsymbol{f}}_{k}, \\
    D_k \tilde{\boldsymbol{u}}_k &= 0, \\
    P(t) \tilde{\boldsymbol{u}}_k &= \widetilde{\boldsymbol{u}}_{\Gamma, k},
    \label{eq:Dis_FT_NS_IB}
\end{aligned}
\end{equation}
where $q$ is a pressure-like variable. In this equation, the Fourier coefficients of the nonlinear term are evaluated with the pseudo-spectral approach discussed above. $P(t)$ is the IB interpolation operator, which is based on the discrete delta function approach \cite{peskin2002immersed}.  Any discrete delta function can be used in the formulation; the calculations below utilized a three-point delta function \cite{roma1999adaptive}
\begin{equation}
    \phi(r) = 
    \begin{cases}
        1 + \sqrt{-3r^2 + 1} , |r| < 0.5, \\
        \frac{1}{6}(5 - 3|r| - \sqrt{1 - 3(1-|r|)^2}), |r| \in [0.5, 1.5),\\
        0 \textrm{ otherwise}.
    \end{cases}
\end{equation}

\section{Lattice Green's functions}
\label{sec:lgf}

\subsection{Lattice Green's Function of $L_k$}

In solving Eq.~\ref{eq:Dis_FT_NS_IB}, substituting the momentum equation in the divergence-free constraint gives rise to an inhomogeneous screened Poisson equation, $L_k u = f$ that must be solved at each time sub-step.  We utilize the lattice Green's function (LGF) on a formally infinite grid to solve this system.  For each $k$, we can find a LGF, $B_k:\mathbb{Z}^2\rightarrow \mathbb{R}$, of the operator $L_k$ such that
\begin{equation}
    (L_k B_k)(n,m) = (\Delta x) ^ 2\delta^{\mathbb{Z}}(n - m), \quad \lim\limits_{n,m \rightarrow \infty} B_k(n,m) = 0 
\end{equation}
where $\delta^{\mathbb{Z}}:\mathbb{Z}\rightarrow\{0,1\}$ is the Kronecker delta function and defined as:
\begin{equation}
    \delta^{\mathbb{Z}}(n) = \begin{cases}
        1 \quad \textrm{if } n = 0, \\
        0 \quad \textrm{if } n \neq 0.
    \end{cases}
\end{equation}
As a result, we can solve the inhomogeneous screened Poisson equation using this LGF \cite{katsura1971lattice, martinsson2002asymptotic}:
\begin{equation}
    L_k u = f \Rightarrow u(n,m) = (L_k^{-1}f)(n,m):=\sum\limits_{a, b}(\Delta x)^2 B_k(n-a, m-b)f(a,b).
\end{equation}
We call the sequence $\{B_k(n,m)\}_{n,m}$ the LGF kernel.

When $k \ne 0$, we can write the LGF kernel as \cite{buneman1971analytic}:
\begin{equation}
    {B}_k(n,m) - {B}_k(0,0) = \frac{1}{2\pi}\int\limits ^{\pi}_{-\pi} \left(1 - \frac{e^{i\theta m}}{K^{|n|}}\right) \frac{d\theta}{K - 1/K}
    \label{eq:integral_LGF}
\end{equation}
where 
\begin{equation}
    K = \frac{a + \sqrt{a^2 - 4}}{2},
\end{equation}
and 
\begin{equation}
    a = \left( 4 + \left(\frac{2\pi k\Delta x}{c} \right)^2 - 2\cos(\theta) \right).
\end{equation}
Finally, 
\begin{equation}
{B}_k(0,0) = \frac{1}{2b}\,_2F_1\left(\frac{1}{2},\frac{1}{2}; 1;\left(\frac{2}{b}\right)^2\right),
\end{equation}
where $b = 2+2\left(\frac{\pi k \Delta x}{c}\right)^2$. We note that the integral in Eq.~\ref{eq:integral_LGF} is increasingly oscillatory with increasing $m$. However, as the integrand is periodic, it can be approximated with spectral convergence using the trapezoidal rule \cite{trefethen2014exponentially}. Further, one can show that the number of quadrature points needed to evaluate this integral at most increases linearly with $m$. However, we do note that the spectral convergence rate of the numerical approximation is dictated by $a$. Specifically, as we show in \ref{ap:TrapConv}, the approximation error of an N-point trapezoidal rule approximation ($\epsilon_N$) is bounded by
\begin{equation}
    |\epsilon_N| \leq \frac{2 M}{e^{\gamma_c N} - 1},
\end{equation}
where $\gamma_c$ is any positive real number satisfying:
\begin{equation}
    \gamma_c < \log \left(1 + \frac{\alpha^2}{2} + \sqrt{\left(1+\frac{\alpha^2}{2}\right)^2 - 1} \right) := \kappa_c, \qquad \alpha = \frac{2\pi k \Delta x}{c},
\end{equation}
and $M$ is
\begin{equation}
    M = \sup\limits_{|\Im(\theta)| < \gamma_c} \left| (1 - \frac{e^{i\theta m}}{K^{|n|}})\frac{1}{K - 1/K} \right|.
\end{equation}

\begin{figure}
    \centering
    \includegraphics[width=0.7\textwidth]{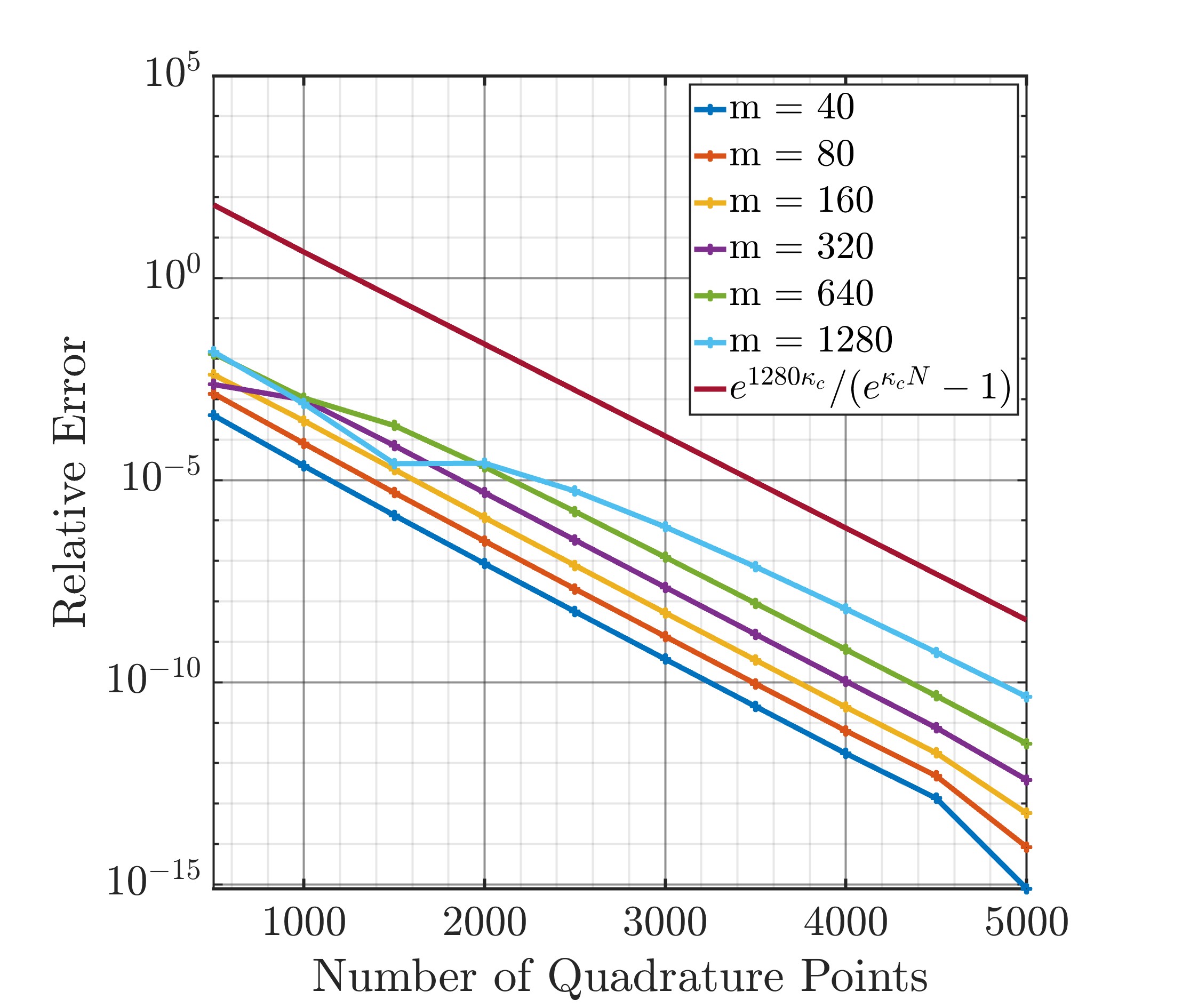}
    \caption{Convergence of $B(n,m)$ for $n = m = 1280$ and various $n$ compared with the analytical convergence rate.}
    \label{fig:LGF_err_spectralBounded}
\end{figure}

In our implementation, this integral is evaluated for each $k$, $m$, and $n$ using the adaptive trapezoidal rule \cite{schaling2011boost} with relative tolerance of $10^{-10}$. This algorithm halves the integration step size (i.e. double the quadrature points) until the tolerance is achieved. Convergence is empirically demonstrated in Figure~\ref{fig:LGF_err_spectralBounded}.

\begin{figure}
    \centering
    \includegraphics[width=0.7\textwidth]{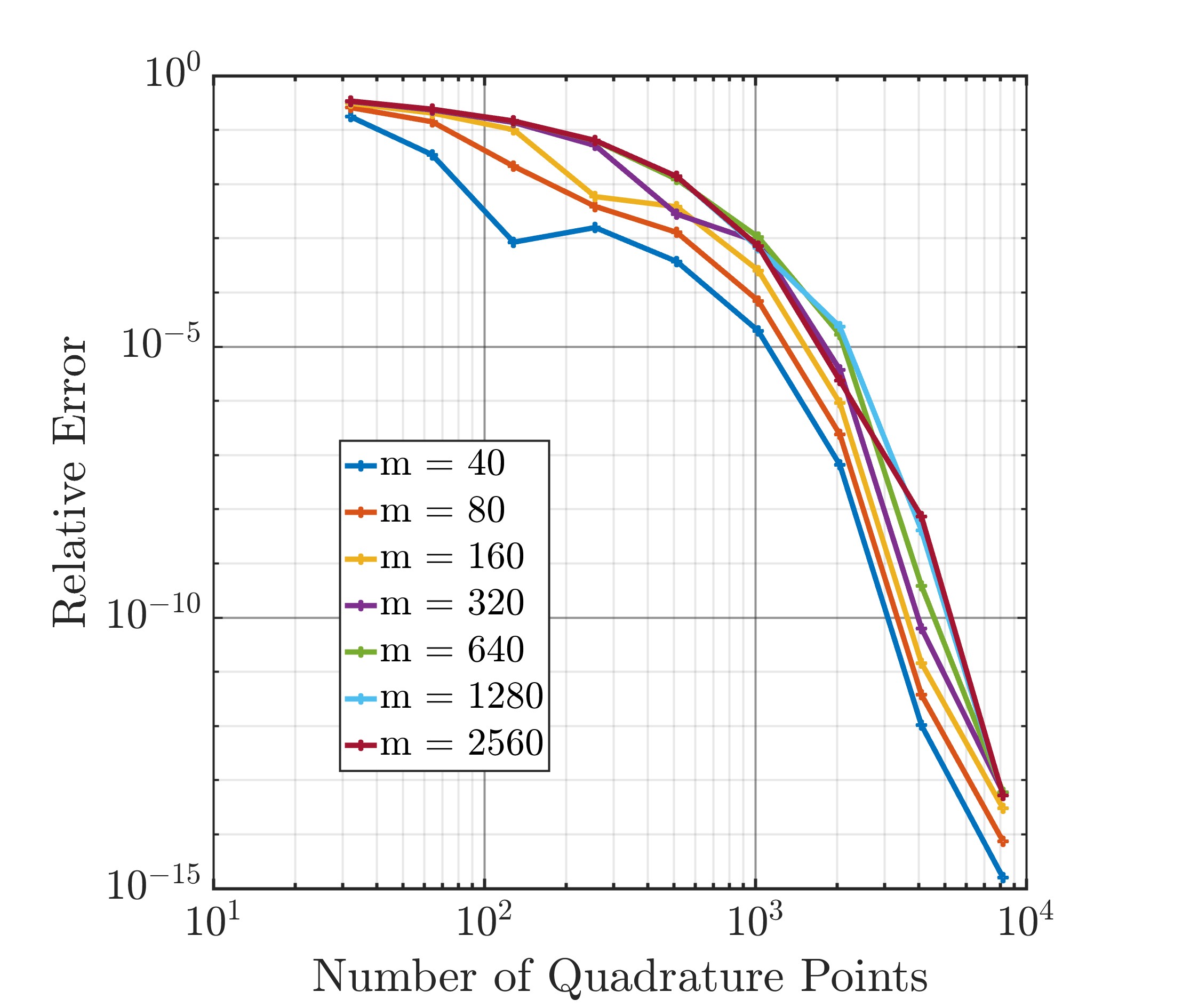}
    \caption{Spectral convergence of $B(m,m)$ for various $m$.} 
    \label{fig:LGF_err_conv}
\end{figure}

\begin{figure}
    \centering
    \includegraphics[width=0.7\textwidth]{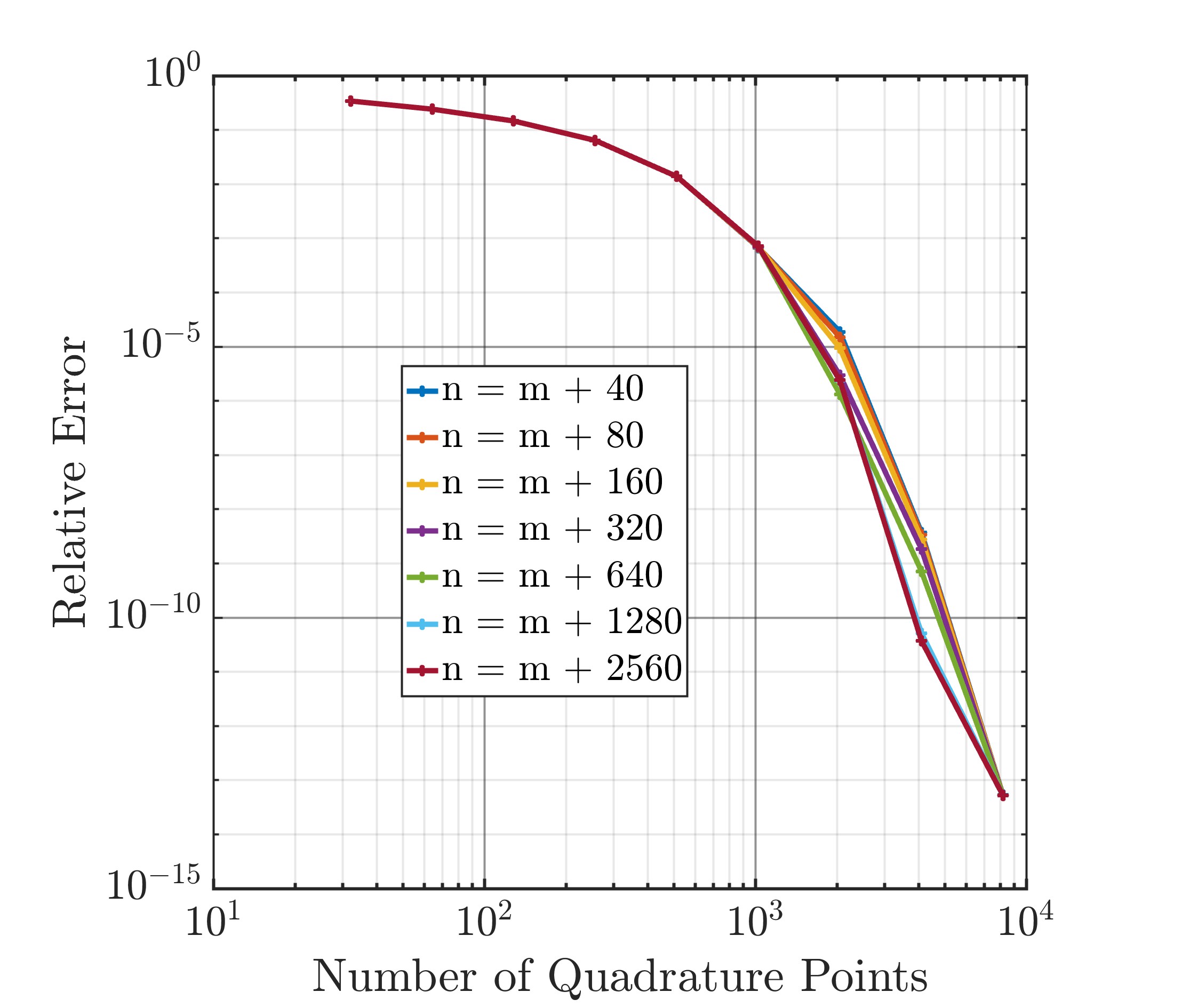}
    \caption{Convergence of $B(n,m)$ for $m = 1280$ and various $n$.}
    \label{fig:LGF_err_conv_var_n}
\end{figure}

  To determine the required number of quadrature points for practical computations, we consider a worst-case scenario. First, consider that $K$ monotonically increases with $\alpha$, which also monotonically increases with $k\Delta x/c$. We thus consider the case where $k = 1$ and $c/\Delta x = 1200$ which provides a conservative estimate for the lowest value of $k\Delta x/c$ likely to be encountered in applications. To further simplify the matter, we observe that $B(n,m) = B(|n|,|m|)$, so, without loss of generality, we can assume $n,m \geq 0$. In addition, since $K > 1$, the higher the value $n$ is, the greater $K^{|n|}$, and the smaller the oscillatory term in the integrand. Thus, the worst case happens when $m$ is large and $n$ is small. However, since $B(m,n) = B(n,m)$ \cite{katsura1971lattice}, we can always write $B(n,m) = B(\max(n,m), \min(n,m))$.  Thus the oscillation of the integrand is the most severe when $m = n$ and they are both large. Computed errors for a range of $m$ are shown in Figure~\ref{fig:LGF_err_conv}. To further illustrate the point that increasing $n$ when holding $m$ constant will not exacerbate the effects of oscillation during numerical integration, we hold $m = 1280$ and vary $n$ in Figure~\ref{fig:LGF_err_conv_var_n}.


Returning now to the case $k = 0$, the far-field boundary condition imposed on $B_k$ is not achievable, since the fundamental solution logarithmically diverges in 2D \cite{martinsson2002asymptotic}. However, eliminating this term provides an inverse that is unique up to a constant.  This constant can be absorbed into the pressure and need never be determined explicitly.  However, when we introduce the multi-resolution LGFs (Section~\ref{sec:apply_lgf_mr}), we shall need to impose a compatibility constraint so that the arbitrary constant is the same regardless of resolution.  Let $B_{0}^l$ denote the kernel on the $l^{th}$ refinement level (0 is the coarsest grid), we can write the compatibility condition as the following:
\begin{equation}
    B_{0}^l(\boldsymbol{n}) = B_0^0(\boldsymbol{n}) - \frac{l}{2\pi}\log(2) \qquad \forall \boldsymbol{n} \in \mathbb{Z}^2.
\end{equation}
We detail the derivation of this compatibility condition in \ref{ap:LGF_compat}.

With a compact source term, $L^{-1}_k$ provides the solution at any point on an infinite lattice.   However, to march the solution (and the source) to the next time, we only need to evaluate the action of applying $L^{-1}_k$ on the support of its source (including a buffer region to allow the solution to adapt).  To ensure the accuracy of the solution and adapt to the evolving vortical flow region, we adopt the domain adaptation and ``velocity refresh'' algorithms developed for the 3D inhomogeneous case \cite{liska2016fast, liska2017fast}. Further details of these techniques will be provided in Section~\ref{sec:adaptivity}. 

To accelerate the application of $L^{-1}_k$, we adopt an interpolation-based kernel-independent fast multipole method on a block-wise decomposed grid \cite{liska2014parallel}. This algorithm not only achieves linear complexity but also lends itself to efficient parallelization.

\subsection{Integrating Factors ($E^i_{k,n}$)}
The availability of the LGF provides an opportunity to use an integrating factor to march the viscous term without an associated time step restriction, enabling the application of an efficient RK-type explicit DAE solver \cite{liska2016fast}.  To implement this in the present method requires finding the LGF for the integrating factor, $H_k(t)$, that solves the following linear ODE system on an infinite lattice:
\begin{equation}
    \frac{dH_k(t)}{dt} = \frac{1}{Re}L_kH_k(t), \quad H_k(0) = I,
    \label{eq:hk}
\end{equation}
where $I$ is the identity operator.

We first denote the kernel of $H_0(t)$ as $A_0(t)$, which can be written as\cite{liska2017fast}: 
\begin{align}
    A_0(t)(\boldsymbol{n}) & = \frac{1}{4\pi^2}\int\limits_{\Pi}\exp\left({-i\boldsymbol{n}.\boldsymbol{\xi} + \frac{\sigma(\boldsymbol{\xi})t}{Re\Delta x^2}}\right)d\boldsymbol{\xi} \nonumber \\ &= \prod\limits_{q \in \boldsymbol{n}}\left[\exp\left(\frac{-2t}{Re\Delta x^2}\right)I_q(\frac{2t}{Re\Delta x^2})\right],
\end{align}
where $\sigma(\boldsymbol{\xi}) = 2\cos(\xi_1) + 2\cos(\xi_2) - 4$, $\Pi = (-\pi, \pi)^2$, and $I_n(z)$ is the modified Bessel function of the first kind of order $n$. For $k \ne 0$, $H_k$ can be found in terms of $H_0$.  The resulting expression is
\begin{equation}
    H_k(t) = \exp\left[-\left(\frac{2\pi k}{c}\right)^2\frac{t}{Re}\right]H_0(t).
\end{equation}
This solution can be verified as follows.  The IC follows by evaluating the expression at $t=0$
\begin{equation}
    H_k(0) = \exp(0)H_0(0) = I,
\end{equation}
and Eq.~(\ref{eq:hk}) follows by
\begin{align}
            \frac{d H_k}{dt} 
            &=\exp\left[-\left(\frac{2\pi k}{c}\right)^2\frac{t}{Re}\right]\frac{dH_0}{dt} - \left(\frac{2\pi k}{c}\right)^2\frac{1}{Re}\exp\left[-\left(\frac{2\pi k}{c}\right)^2\frac{t}{Re}\right]H_0 \\
            &= \exp\left[-\left(\frac{2\pi k}{c}\right)^2\frac{t}{Re}\right]\frac{1}{Re}L_0 H_0 - \left(\frac{2\pi k}{c}\right)^2\frac{1}{Re}H_k \\
            &= \frac{1}{Re}L_0 H_k - \left(\frac{2\pi k}{c}\right)^2\frac{1}{Re}H_k \\
            &= \frac{1}{Re}\left[L_0 - \left(\frac{2\pi k}{c}\right)^2I\right]H_k, \\
        & \equiv \frac{1}{Re} L_k H_k.  
\end{align}

Note that the kernel associated with $H_k(t)$ decays faster than any exponential, which can be exploited in the fast multipole solution by restricting the source of any target to its vicinity \cite{liska2016fast}.

\section{Temporal Discretization}
\label{sec:time_disc}

To discretize Eq.~\ref{eq:Dis_FT_NS_IB} in time, by imposing the integrating factor, we can rewrite the system by denoting $\tilde{\boldsymbol{v}}_k = H_k(t)\tilde{\boldsymbol{u}}_k$ and $\tilde{{b}}_k = H_k(t)\tilde{q}_k$, so we have:
\begin{equation}
    \begin{aligned} 
\frac{d {\tilde{\boldsymbol{v}}}_k}{d t} + H_k(t)\mathcal{F}_k[\boldsymbol{N}(\boldsymbol{\omega}, \boldsymbol{u})] &= -G_k \tilde{b}_k + H_k(t) P(t)^T \tilde{\boldsymbol{f}}_{k}, \\
    D_k \tilde{\boldsymbol{v}}_k &= 0, \\
    P(t) H_k^{-1}(t)\tilde{\boldsymbol{v}}_k &= \widetilde{\boldsymbol{u}}_{\Gamma, k}.
\end{aligned}
\end{equation}
We adopt a half-explicit Runge-Kutta (IF-HERK) method for these DAE \cite{brasey1993half, liska2017fast}.  For the present method, this can be stated as 
\begin{enumerate}
    \item Initialize: set $\tilde{\boldsymbol{u}}^0_{k,n} = \tilde{\boldsymbol{u}}_{k,n}$ and $t_n^0 = t_n$
    \item Multistage: for $i = 1,2, ..., s$, solve the linear system
    \begin{equation}
        \begin{bmatrix}
    (E^i_{k})^{-1} & (Q^{(i-1)}_{k,n})^* \\ Q^{i}_{k,n} & 0 
    \end{bmatrix}
    \begin{bmatrix}
    \tilde{\boldsymbol{u}}^i_{k,n} \\ \hat{\lambda}^i_{k,n}
    \end{bmatrix}
    = \begin{bmatrix}
    \boldsymbol{r}^i_{k,n} \\ \zeta^i_{k,n}
    \end{bmatrix}. \label{eqn:linsys}
    \end{equation}
    \item Finalize: set $\tilde{\boldsymbol{u}}_{k,n+1} = \tilde{\boldsymbol{u}}_{k,n}^s$, $\lambda_{k, n+1} = (\tilde{a}_{s,s}\Delta t)^{-1}\hat{\lambda}^s_{k,n}$, and $t_{n+1} = t_n^s$
\end{enumerate}
where 
\begin{equation}
Q_{k,n}^i=
    \begin{bmatrix}
    G_k^* \\ P^i_n
    \end{bmatrix}
    , \qquad 
    \hat{\lambda}^i_{k,n} =
    \begin{bmatrix}
        \tilde{q}_{k,n}^i \\ \tilde{f}^i_{k,n}
    \end{bmatrix}
    , \qquad \zeta^i_{k,n} =
    \begin{bmatrix}
        0 \\ (\widetilde{\boldsymbol{u}}_{\Gamma, k})^i_{n}
    \end{bmatrix}.
\end{equation}

The terms appearing in the linear system are
    \begin{equation}
        \boldsymbol{u}_n^{i-1} = \mathcal{F}^{-1}(\{\tilde{\boldsymbol{u}}^{i - 1}_{k,n}\}_{k}), \qquad \boldsymbol{\omega}_n^{i-1} = \mathcal{F}^{-1}(\{C_k\tilde{\boldsymbol{u}}^{i - 1}_{k,n}\}_{k})
    \end{equation}
    where
    $\mathcal{F}^{-1}$ is evaluated using (de-aliased) inverse Fast Fourier Transform (iFFT).  Furthermore
    \begin{equation}
        E^i_k = H_k\left(\frac{(\Tilde{c}_i - \Tilde{c}_{i-1})\Delta t}{(\Delta x)^2Re}\right), \qquad g^i_{k,n} = -\Tilde{a}_{i,i}\Delta t \mathcal{F}_k[\boldsymbol{N}(\boldsymbol{\omega}_n^{i-1}, \boldsymbol{u}_n^{i-1})]
        \label{eq:EK_RK}
    \end{equation}
    where for all $k$, $\mathcal{F}_k[\boldsymbol{N}(\boldsymbol{\omega}_n^{i-1}, \boldsymbol{u}_n^{i-1})]$ is evaluated using FFT. Then the following are recursively computed for $i > 1$ and $j > i$ using:
    \begin{equation}
        h^i_{k,n}=E^{i-1}_kh^{i-1}_{k,n}, \qquad h^1_{k,n} = \tilde{\boldsymbol{u}}^0_{k,n}
    \end{equation}
    \begin{equation}
        w^{i,j}_{k,n} = E^{i - 1}_kw^{i-1, j}_{k,n}, \qquad w^{i,i}_{k,n} = (\Tilde{a}_{i,i}\Delta t)^{-1}\left(g^{i}_{k,n} - Q^{i-1}_{k,n}\hat{\lambda}^{i}_{k,n}\right).
    \end{equation}
    And we compute:
    \begin{equation}
        r^i_{k,n} = h^i_{k,n} + g^i_{k,n} + \Delta t \sum\limits_{j = 1}^{i-1}\Tilde{a}_{i,j}w^{i,j}_{k,n} , \qquad t^i_n = t_n + \Tilde{c}_i\Delta t .
        \label{eq:g_kt_k}
    \end{equation}

With this time discretization, at the $n^{th}$ time step in $i^{th}$ stage for the $k^{th}$ Fourier coefficient, we have the following linear system:
\begin{equation}
        M^i_{k,n} 
    \begin{bmatrix}
    \tilde{\boldsymbol{u}}^i_{k,n} \\ \tilde{q}^i_{k,n} \\ \tilde{f}^i_{k,n}
    \end{bmatrix}
    = \begin{bmatrix}
    (E^i_{k})^{-1} & G_k & (P^{(i-1)}_n)^T \\ G_k^* & 0 & 0 \\ P^{i}_n & 0 & 0
    \end{bmatrix}
    \begin{bmatrix}
    \tilde{\boldsymbol{u}}^i_{k,n} \\ \tilde{q}^i_{k,n} \\ \tilde{f}^i_{k,n}
    \end{bmatrix}
    = \begin{bmatrix}
    \boldsymbol{r}^i_{k,n} \\ 0 \\ (\widetilde{\boldsymbol{u}}_{\Gamma, k})^i_{n}
    \end{bmatrix}
    \label{eq:linsys_RK3}
\end{equation}
where $S^i_{k,n} = P^i_{n}E^i_{k}(I - G_kL_k^{-1}D_k)(P^{i-1}_n)^T$, which we will show, in \ref{ap:S_hermitian}, that it is Hermitian when $P^i_{n} = P^{i-1}_n$. $E^i_{k}$ refers to the integrating factor associated with $L_k$. $L^{-1}_k$ refers to the lattice Green's Function (LGF) of $L_k$. We will explain how to apply them in subsequent sections. 

We can solve the linear system arising from the IF-HERK method using a block LU decomposition. As a result, the steps to solve this system are:
\begin{equation}
    \begin{aligned}
    \tilde{q}^*_k &= -L_k^{-1}G^*_k\boldsymbol{r}_{k,n}^i, \\
    S^i_{k,n}\tilde{f}^i_{k,n} &= P^i_kE^i_{k}[\boldsymbol{r}_{k,n}^i - G_k\tilde{q}^*_k] - (\widetilde{\boldsymbol{u}}_{\Gamma, k})^i_{n}, \\
    \tilde{q}_{k,n}^i &= \tilde{q}^*_k + L_k^{-1}G^*_k(P^i_n)^T\tilde{f}^i_{k,n}, \\
    \tilde{\boldsymbol{u}}_{k,n}^i &= E^i_{k}[\boldsymbol{r}_{k,n}^i - G_k\tilde{q}_{k,n}^i - (P^{i-1}_n)^T\tilde{f}^i_{k,n}].
\end{aligned}
\label{eq:block_LU}
\end{equation}
Note that this block LU decomposition method does not have splitting error due to the use of the integrating factor \cite{liska2017fast}. The advantage is that the divergence-free constraint and the boundary conditions are satisfied exactly \cite{taira2007immersed, liska2017fast}. In our current implementation of this algorithm, we use a 3rd order scheme with the Butcher Tableau shown in Table~\ref{tb:rk3}. 
\begin{table}[]
\center
\begin{tabular}{l|lll}
0   & 0   & 0   & 0   \\
1/3 & 1/3 & 0   & 0   \\
1   & -1  & 2   & 0   \\ \hline
    & 0   & 3/4 & 1/4
\end{tabular}
\caption{Runge-Kutta scheme Butcher Tableau used in our implementation}
\label{tb:rk3}
\end{table}

Apart from the second equation of Eq.~\ref{eq:block_LU}, the remaining equations can be solved directly through the application of LGF and the integrating factor. The second equation in Eq.~\ref{eq:block_LU} corresponds to the projection step to compute the IB forcing in order to satisfy the boundary condition \cite{taira2007immersed,liska2017fast}. In similar algorithms designed for general 3D flows \cite{liska2017fast, yu2022multi}, the conjugate gradient method is employed to solve for the IB forcing. In that case, it is estimated that the linear system associated with IB forcing is a $3N_{L}$ by $3N_L$ dense system where $N_L$ is the total number of IB points, and the constant $3$ arises from the three velocity components. In the case of general 3D flows, the number of operations needed to solve such a linear system is $O(N_L^3)$ \cite{liska2017fast}. In addition, due to the sequential nature of matrix factorization and back-substitution \cite{liska2017fast}, directly solving for IB forcing using numerical factorization becomes less desirable than the conjugate gradient method. 

For flows with one homogeneous direction, the immersed body has a uniform 2D cross-section across the span-wise direction giving $N_L = N_z N_{2D}$, where $N_z$ is the number of Fourier coefficients in the truncated Fourier series, and $N_{2D}$ is the number of IB points used to represent the 2D cross-section. To solve for the Fourier coefficients of the IB forcing using the evolution equations of the Fourier coefficients (Eq.~\ref{eq:FT_NS_ib}), we solve for $N_z$ independent $3N_{2D}$ by $3N_{2D}$ dense linear systems instead of one $3N_L$ by $3N_L$ dense linear system. Thus, the operation count for using a direct solver in our scenarios decreases to $O(N_z (N_{2D})^3) = O(N_L^3/N_z^2)$. More importantly, due to the independence among those $N_z$ linear systems, the application of numerical factorization and back-substitution can be efficiently parallelized. Thus, unlike the method to solve for IB forcing in the general 3D flow solver algorithm, we solve the IB forcing using direct LU factorization instead of the conjugate gradient method. In our implementation, the dense linear system is solved using ScaLAPACK\cite{blackford1997scalapack} wrapper from PETSc\cite{petsc-user-ref}. According to our numerical experiments, solving for IB forcing directly takes less than $3\%$ of the total computational time when using the LU factorization.

\section{Multi-resolution mesh}
\label{sec:apply_lgf_mr}

To efficiently resolve thin boundary layers, particularly with the IB method, adaptive mesh refinement is needed \cite{mittal1997inclusion, lehmkuhl2013low, yu2021multi}. For Cartesian grids, a wide range of adaptive mesh refinement (AMR) algorithms have been proposed to locally refine the computational mesh \cite{berger1984adaptive, berger1989local, offermans2020adaptive}. If we restrict our attention to one 2D slice of the computational grid in our algorithm, the existing AMR algorithm applies. However, we require that the AMR algorithm be compatible with the application of LGF. For this reason, on each of the 2D slices, we employ an existing adaptive mesh refinement approach that has proven efficient and accurate when combined with LGF \cite{yu2021multi}. This AMR approach can be directly applied to each 2D slice in the $x-y$ plane for each Fourier coefficient. Adaptation of this AMR algorithm to our hybrid method is detailed in the remainder of this section.

\subsection{Multi-resolution mesh on a Cartesian grid}

We first review the salient features of the algorithm of Yu et al. \cite{yu2021multi}. We start by constructing a composite grid consisting of multiple computational grids with increasing resolutions, $\{\mathbb{R}^{\mathcal{Q}}_0, \mathbb{R}^{\mathcal{Q}}_1, ..., \mathbb{R}^{\mathcal{Q}}_{N}\}$, wherein $\mathbb{R}^{\mathcal{Q}}_k$ is generated by evenly dividing each computational cell in $\mathbb{R}^{\mathcal{Q}}_{k - 1}$ into $2^d$ smaller cells, $d$ being the physical dimension of the problem. The composite grid is the Cartesian product of this series of grids:
\begin{equation}
    \overline{\mathbb{R}^{\mathcal{Q}}} = \bigotimes_{k = 0}^{N} \mathbb{R}^{\mathcal{Q}}_k.
\end{equation}
Then the AMR grid is defined by partitioning the entire computational domain $\Omega$ into $N + 1$ smaller pairwise-disjoint domains $\{\Omega_0, \Omega_1, ... \Omega_{N}\}$. Define the restriction functional as:
\begin{equation}
    \Gamma_k(f)(\boldsymbol{n}) =\begin{cases}
    f(\boldsymbol{n}) &\textrm{ if } \boldsymbol{n} \in \Omega_k,\\
    0 &\textrm{ if } \boldsymbol{n} \notin \Omega_k
    \end{cases},
\end{equation}
and the values on the AMR computational grid are defined through the Cartesian product of these restriction functionals
\begin{equation}
    \overline{\Gamma} = \bigotimes_{k = 0}^{N}\Gamma_k.
\end{equation}
The operator $\overline{\Gamma}$ restricts the region we need to compute the numerical solutions to
\begin{equation}
    \overline{\Omega} = \bigotimes_{k = 0}^{N}\mathbb{R}^{\mathcal{Q}}_k \cap \Omega_k.
\end{equation}
Thus, the solution is defined in the subspace $\overline{\Omega}$. To communicate the information across different levels of mesh, we also define interpolation and coarsening operators from level $l$ to level $k$ as $P_{l\rightarrow k}$ (interpolation when $l < k$ and coarsening when $l > k$). To estimate the information $f$ on level $k$ given the information across the AMR grid $\hat{f}$, we compute:
\begin{equation}
    f_k = \sum\limits_{l = 0}^{N}P_{l \rightarrow k}\hat{f}_l.
\end{equation}

Then to apply LGF on the AMR mesh from a source term $\hat{f}$, we use the following \cite{yu2021multi}:
\begin{equation}
    \phi(\boldsymbol{n}) = \otimes_{j = 0}^{N} \left[ \left(\sum\limits_{i=0}^{j-1} \overline{P}^Q_{i\rightarrow j}L^{-1}_{0,i} \hat{f}_i \right) + L^{-1}_{0,j} \left(  \sum\limits_{i=j}^{N} P^Q_{i\rightarrow j} \hat{f}_i \right) \right],
    \label{eq:apply_lgf_2D}
\end{equation}
where $\overline{P}^Q_{i\rightarrow j}$ is a commutative projection operator (see Eq.30-32 in \cite{yu2022multi}). We use $L^{-1}_{0,l}$ to denote the action of applying LGF of Laplacian on level $l$ in the refinement mesh. In this equation, at level $j$, the first term recursively computes the solution at level $j$ induced from the solution at coarser levels. The second term computes the solution induced by the source on level $j$ and the source interpolated from finer levels. The first term is accumulated when computing the solution from the base level to the finest level. We can directly use this method to apply $L_k^{-1}$ for each Fourier coefficient. To achieve additional speedup when applying LGFs, a fast and parallel multi-resolution elliptic equation solution method is employed \cite{dorschner2020fast, ying2004kernel, liska2014parallel}.

\subsection{Multi-resolution in Fourier space}

To efficiently exploit the multi-resolution mesh in the $x-y$ plane, we must also locally truncate the Fourier series such that the resolution in $z$ is comparable to the local resolution in the $x-y$ plane.  Considering the spectral convergence in $z$, compared to the low-order convergence of the finite volume discretization in the $x-y$ plane, it is expected that the mesh in the $z$ direction needs to be refined, at most, as fast as the rate we refine the mesh in the $x-y$ plane.

However, for spanwise homogeneous flows, the flow field is not homogeneous everywhere in the presence of the immersed body, especially when the boundary layer is laminar. In the far wake, the length scales tend toward homogeneity, but, near the immersed body, variation in the $x-y$ plane can be much more rapid than that of the $z$-direction \cite{smith1986steady, williamson1996vortex}. One example of such inhomogeneity is the flow past a circular cylinder. In the shear-layer transition regime ($Re \sim 1,000-200,000$), two shear layers are forming from the side of the cylinder. Those shear layers and the associating Kelvin-Helmholtz instability are essentially 2D. Thus, to resolve the shear layers, we only need to refine the computational grid in the $x-y$ plane. However, downstream of the shear layers, the flow transitions to three-dimensional turbulence, and comparable resolution is required in all three directions.  %
To optimally treat these situations, we modify the spanwise refinement. Far from the body, if there are $l_{ref}$ refinement levels, on level $l$, we retain $N_l = N_0 \times 2^l$ Fourier coefficients, where $N_0$ is the number for the base (coarsest) mesh.  Near the body, we cap the number of Fourier coefficients, even as we refine the $x-y$ grid by an additional $l_{add}$ levels.

We now elaborate on how we apply the LGF on the multi-resolution mesh in both Fourier space and the $x-y$ plane. For the $k^{th}$ Fourier coefficient, we find an $l$ such that $N_0 2^{l-1}\leq k < N_0 2^l$. We know that computational cells on level $l-1$ need only retain the first $N_02^{l-1}$ Fourier coefficients. That is, we may assume the $k^{th}$ Fourier coefficient is zero for all computational cells on level $l - 1$. Similarly, the $k^{th}$ Fourier coefficient is zero for all computational cells on levels 0 through $l - 1$. Thus, the $k^{th}$ Fourier coefficients and the associating source terms $\hat{f}^k$ are zero on those levels. We do not need to apply $L_k^{-1}$ on levels 0 through $l - 1$, nor need we consider source terms from those levels since they are to be truncated. With this strategy, we can simplify the procedure that applies $L_k^{-1}$ in Eq.~\ref{eq:apply_lgf_2D} so that the index of the first term begins at $i=l$ rather than $i=0$, resulting in significant computational savings.

Conversely, for the $k^{th}$ Fourier coefficient, the corresponding coarsest level that $L_k^{-1}$ need be applied is given by:
\begin{equation}
    l = \max\left(\left\lceil\log_2\left(\frac{k+1}{N_0}\right)\right\rceil, 0 \right) = \max\left(l_{ref} - \left\lfloor\log_2\left(\frac{N_0\times 2^{l_{ref}}}{k+1}\right)\right\rfloor, 0 \right).
    \label{eq:coarse_nonzero_level}
\end{equation}

Similar to applying $L^{-1}_k$, applying discrete operators (divergence, gradient, curl), interpolation operators, and the integrating factor $E^i_{k}$ follows a similar strategy. For an operator operating on the $k^{th}$ Fourier coefficient, the coarsest level, $l$, the operator needs to be applied on is also determined by Eq.~\ref{eq:coarse_nonzero_level}. Thus. that operator only needs to be applied to grid points on levels greater than or equal to $l$.

\section{Adaptation}
\label{sec:adaptivity}

So far, we have introduced the steps to time integrate the discretized Navier-Stokes equations using the IF-HERK method and LGF. In addition, the computational grid spatially adapts to the vortical region, which we term base-level adaptivity, and adaptively refines the mesh in a blockwise fashion.  For adaptation, we adopt the strategies developed for the fully inhomogeneous case \cite{liska2016fast,yu2021multi}; we provide a brief summary of the algorithms here. 

\subsection{Base-level (domain) adaptation}

The assumption that allows us to constrain our computational domain to a finite one is that the vortical region is in general compact (exponentially decaying). 
 The strategy is to add or delete cells (blockwise) when the vorticity near the boundary exceeds or falls below a threshold value.  However, the surface defining the threshold value must adapt in time as the vorticity evolves in the boundary layer and wake regions.  Additionally, it is sometimes pertinent to alter the threshold in the far wake as, for long times, the vorticity decays slowly.
 
 To these ends, we denote the active cell region in the base level
\begin{equation}
    \Omega_0^{supp} =\biggl\{ \boldsymbol{x} \in \mathbb{R}^2: \frac{ \sum_i |D_i\boldsymbol{r}_{i}(\boldsymbol{x})|^2 }{\max\limits_{\boldsymbol{x}}\sum_i |D_i\boldsymbol{r}_{i}(\boldsymbol{x})|^2 } \geq \epsilon_0 \biggl\},
    \label{eq:base_adapt_criterion}
\end{equation}
where $\boldsymbol{r}_{i}$ is the nonlinear term (Lamb vector) in the time-discretized equations (Eq.~\ref{eq:g_kt_k}) and $D_k$ the discrete divergence operator for the $k^{th}$ Fourier coefficient. $\epsilon_0$ is some prescribed cutoff threshold. The $\sum_i$ is the sum across all Fourier coefficients. Thus, the term $\sum_i |D_i\boldsymbol{r}_{i}(\boldsymbol{x})|^2$ is approximatively the magnitude of the source term for the pressure-Poisson equation at location $\boldsymbol{x}$ across all Fourier coefficients. To ensure that $\Omega_0^{supp}$ is captured by our computational domain, we additionally put a region ($\Omega_{\textrm{buff}}$) of buffering computational cells around $\Omega_0^{supp}$. These additional computation cells are those at a fixed distance $N_{b}$ from $\partial\Omega_0^{supp}$. We periodically update $\Omega_0^{supp}$ by incorporating grid cells from $\Omega_{\textrm{buff}}$ to satisfy Eq.~\ref{eq:base_adapt_criterion}. The detail of how to choose $N_b$ is discussed in \cite{liska2016fast}.

Each time we incorporate computational cells from $\Omega_{\textrm{buff}}$ into $\Omega_0^{supp}$, a new 
$\partial\Omega_0^{supp}$ is effectively defined, and we need to compute the velocity field in the newly incorporated region $\Omega_{\textrm{buff}}$. To do so, we solve the vorticity-streamfunction equation to fill in the velocity in $\Omega_{\textrm{buff}}$. Let $\Tilde{\boldsymbol{u}}^{u}_k$ be the $k^{th}$ Fourier coefficient of velocity before solving the vorticity-streamfunction equation and $\Tilde{\boldsymbol{u}}^{a}_k$ be the values after, we solve the velocity within $\Omega_{\textrm{buff}}$ using:
\begin{equation}
\begin{aligned}
     &\Tilde{\boldsymbol{\omega}}_k = C_k \Tilde{\boldsymbol{u}}^{u}_k, \\
     &\Tilde{\boldsymbol{u}}^a_k = -C_k^*L_k^{-1}\mathds{1}_{\Omega_0^{supp}}\Tilde{\boldsymbol{\omega}}_k.
     \label{eq:pad_vel}
\end{aligned}
\end{equation}
where $\mathds{1}_{\Omega_0^{supp}}$ is the indicator function of the set $\Omega_0^{supp}$. Namely, when fill in the velocity in the buffering region, we only use the vorticity within $\Omega_0^{supp}$. We term the above procedure as velocity padding.

\subsection{Velocity Refresh}
When applying the integrating factor $E_k^i$ through convolution, the support of the associating kernel is unbounded. However, we can still accurately evaluate this convolution with $\Omega_0^{supp}$ using only the values within 
$\Omega_0^{supp}$ and $\Omega_{\textrm{buff}}$. To that end, we use two facts: the integrating factor decays faster than any exponential \cite{liska2016fast}, and velocity in $\Omega_0^{supp}$ can be obtained by solving a vorticity-streamfunction equation. Thus, we employ a two-step approximation to evaluate the action of $E_k^i$ within $\Omega_0^{supp}$.

First, we truncate the integrating factor kernel to have a compact support by thresholding the value of the kernel. Due to the fast decay, we can very accurately approximate the action of $E_k^i$ by only applying the integrating factor kernel within this compact region. 

Second, near $\partial\Omega_0^{supp}$, the source region of the approximated $E_k^i$ operator extends outside of $\Omega_0^{supp}$. We assume that this additional source region is contained in some $\Omega_{\textrm{IF}}$. Since this region is outside of $\Omega_0^{supp}$, the vorticity within $\Omega_{\textrm{IF}}$ is negligible. Thus, we can compute the velocity in this region using vorticity-streamfunction equation as in Eq.~\ref{eq:pad_vel}. This step is called velocity refresh.

Combining these two steps, we can accurately approximate the action of $E_k^i$ within $\Omega_0^{supp}$ by carefully defining $\Omega_{\textrm{IF}}$ and the source region of $E_k^i$. The specific procedure of properly truncating the integrating factor kernel and defining $\Omega_{\textrm{IF}}$ can be found in \cite{liska2016fast}. We note that the vorticity-streamfunction equation only needs to be solved periodically to ensure accurate simulation. In fact, the frequency we need to solve the vorticity-streamfunction equation depends on $\Omega_{\textrm{IF}}$. 

It is shown that we can overlap $\Omega_{\textrm{IF}}$ with the buffering region we defined for velocity padding \cite{liska2016fast}. That is $\Omega_{\textrm{IF}}$ = $\Omega_{\textrm{buff}}$. As a result, we can compute a corresponding maximum time step $n_r$ such that we only need to conduct velocity refresh every $n_r$ time step. 

\subsection{Adaptive Refinement}

As the flow develops, the high vorticity and high velocity gradient regions change. As a result, the computational grid needs to adapt to the evolution of the flow. In our algorithm, this step is done by tracking the high vorticity region and locally refining the mesh accordingly. Recall that the AMR grid decomposes the entire computational domain into a sequence of pairwise disjoint domains $\{\Omega_l\}_l$. To adaptively refine the computational mesh, we partition each $\Omega_l$ into smaller computational blocks. As shown in Section~\ref{sec:apply_lgf_mr}, the adaptative refinement in the spanwise direction is determined by the adaptive refinement in the $x-y$ plane. Thus, it is sufficient to consider the refinement criterion for the 2D plane formed by the inhomogeneous directions. In this part, the vorticity is used as the criterion for adaptation. We specify a refinement factor $\alpha \in (0, 1)$ and a deletion factor $\beta\in (0, 1)$. When a computational block is on level $l$, we refine the block if any point $\boldsymbol{x}$ in that block satisfies:
\begin{equation}
    S(\boldsymbol{x}) = \sum_k ||\Tilde{\boldsymbol{\omega}}_k(\boldsymbol{x})||_2^2 > \alpha^{l_{max} - l}S_{max},
\end{equation}
and coarsen the block if every point $\boldsymbol{x}$ in that block satisfies:
\begin{equation}
    S(\boldsymbol{x}) = \sum_k ||\Tilde{\boldsymbol{\omega}}_k(\boldsymbol{x})||_2^2 < \beta\alpha^{l_{max} - l}S_{max},
\end{equation}
where $||\Tilde{\boldsymbol{\omega}}_k(\boldsymbol{x})||_2$ is the 2-norm of the $k^{th}$ Fourier coefficient of vorticity vector at point $\boldsymbol{x}$. Using Parseval's identity, we have:
\begin{equation}
\begin{aligned}
  &\frac{1}{c}\int\limits_{-c/2}^{c/2}\sum\limits_{i \in \{1,2,3\}}\omega_i^2(x,y,z)dz\\
= &\sum\limits_{k = -\infty}^{\infty} ||\Tilde{\boldsymbol{\omega}}_k(\boldsymbol{x})||^2 \\
\approx &\sum\limits_{k = -N/2}^{N/2} ||\Tilde{\boldsymbol{\omega}}_k(\boldsymbol{x})||^2 .
\end{aligned}
\end{equation}
Thus, $S(\boldsymbol{x})$ approximates the squared $L_2$ norm of the vorticity at each $\boldsymbol{x}$ location. In addition, we solve the vorticity-streamfunction equations to pad velocity when new blocks are refined using Eq.~\ref{eq:pad_vel}. This is to fill in the Fourier coefficients of velocities in the newly refined blocks, as those Fourier coefficients were previously set to zero due to the truncation of the Fourier series from the multi-resolution nature in Fourier space.

\section{Algorithm Summary}
\label{sec:alg_sum}

Algorithm~\ref{alg:timemarch} summarizes the required steps to march the solution forward by $N$ time steps.  Let $n_a$ be the desired frequency (number of steps) to adapt the domain and/or resolution, and let $n_r$ be the desired frequency (number of steps) to conduct velocity refresh. When simulating fluid flows with this algorithm, $n_a$ should be chosen according to the resolution requirements, and $n_r$ should be chosen according to the procedure detailed in \cite{liska2017fast}.

\begin{algorithm}
\caption{Time Marching using IF-HERK, IB, and LGF}
\begin{algorithmic}[1]

\Procedure{Time Marching}{$\tilde{\boldsymbol{u}}_{k,0}, t_f$}
\State $n = 0$
\While{$n < N$}
    \If{$n \% n_a$ = 0}
    \State Perform domain adaptation
    \State Perform velocity padding using Eq.~\ref{eq:pad_vel}
    \ElsIf{$n \% n_r$ = 0}
    \State Perform velocity refresh using Eq.~\ref{eq:pad_vel}
    \EndIf
\State set $\tilde{\boldsymbol{u}}^0_{k,n} = \tilde{\boldsymbol{u}}_{k,n}$ and $t_n^0 = t_n$
\For{each stage $i \in \{1,2,3\}$}
    \State Compute $\boldsymbol{u}_n^{i-1}$ and $\boldsymbol{\omega}_n^{i-1}$ using inverse FFT
    \State Compute $g_{k,n}^i$, $r_{k,n}^i$ according to Eq.~\ref{eq:EK_RK} and Eq.~\ref{eq:g_kt_k}
    \State Solve the system of equations shown by Eq.~\ref{eq:linsys_RK3} using the block LU decomposition as detailed in Eq.~\ref{eq:block_LU}
\EndFor

\State Setting $\tilde{\boldsymbol{u}}_{k,n+1} = \tilde{\boldsymbol{u}}_{k,n}^3$, $\lambda_{k, n+1} = (\tilde{a}_{3,3}\Delta t)^{-1}\hat{\lambda}^3_{k,n}$, and $t_{n+1} = t_n^3$
\State $n = n+1$

\EndWhile
\EndProcedure
\end{algorithmic}
\label{alg:timemarch}
\end{algorithm}

\section{Parallelization and performance}
\label{sec:parallel}

We adopted a server-client model for parallelization based on decomposing the domain into pencils that correspond to blocks in the $x-y$ plane.  That is, all Fourier coefficients (regardless of the number) are stored on the same processor, which avoids data transfer to accomplish the FFT.  Each block is assigned a computational load according to their roles during the time-stepping routine, and a load balancing algorithm distributes those blocks into different processors \cite{yu2022multi}.

However, since direct solvers are used to solve for IB forcings, we need to devise a corresponding parallelization strategy. Suppose we are solving systems with $N$ Fourier coefficients and $M$ parallel client processes. Two separate parallelization strategies are devised for the case when $N > M$ and $ N \leq M$, respectively.
\begin{itemize}
    \item When $N > M$, each linear system is only solved using one process, and each process is tasked with solving one or more linear systems. Specifically, $N \% M$ processes are allocated to handle $\lceil N/M \rceil$ dense linear systems, and the rest processes are allocated to handle $\lfloor N/M \rfloor$ dense linear systems.
    \item When $N \leq M$, multiple processes are allocated to solve one linear system, and each process is assigned only one linear system. Specifically, $M \% N$ linear systems are each solved by a group of $\lceil M / N \rceil$ processes, while the rest linear systems are each solved by a group of $\lfloor M/N \rfloor$ processes.
\end{itemize}

\begin{figure}
   \centering
   \includegraphics[width=\textwidth]{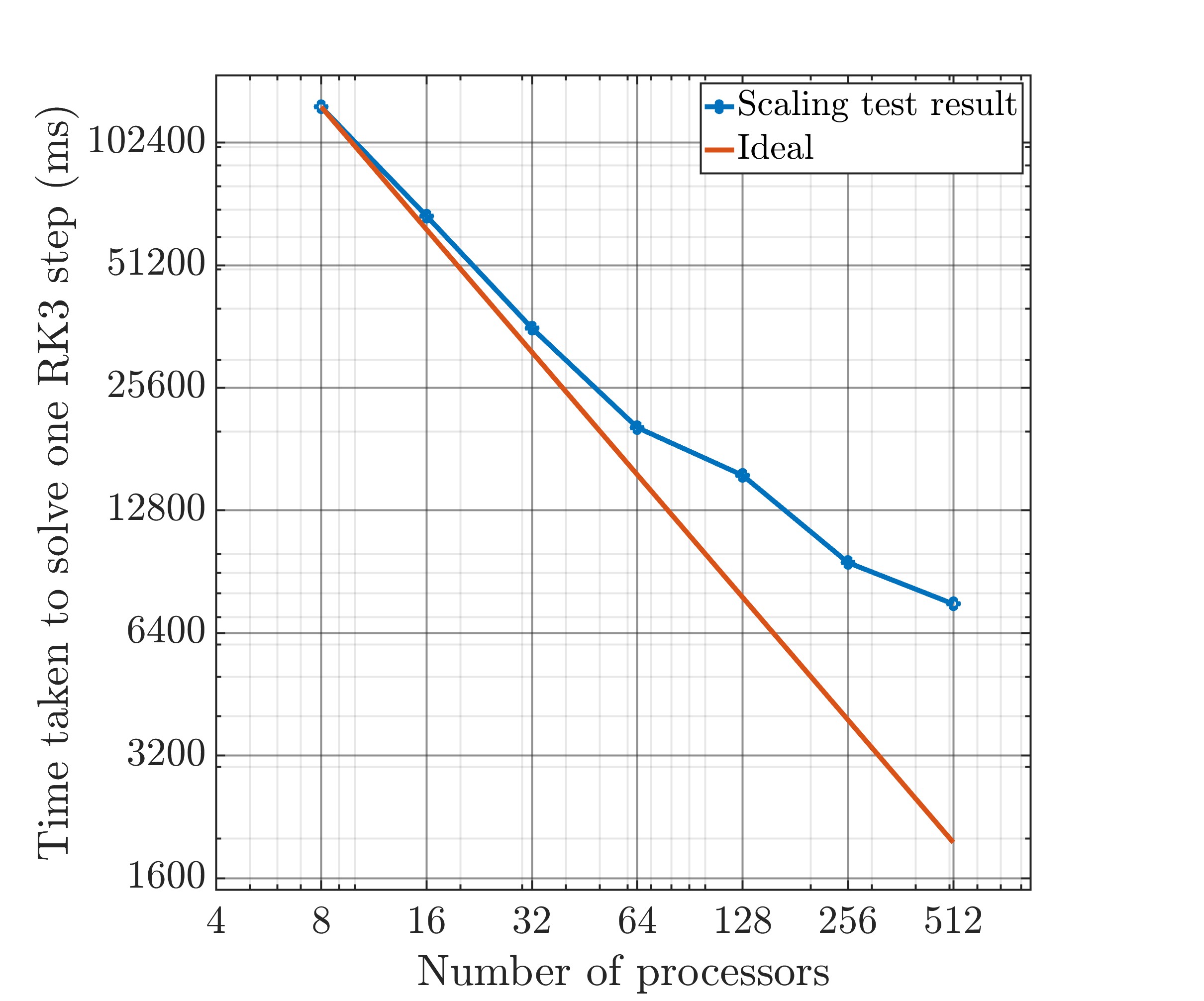}
   \caption{Time to solve one RK3 step vs. the number of cores used in the simulation. The simulations are conducted on PSC Bridges-2 supercomputer \cite{brown2021bridges} with AMD EPYC 7742 2.25GHz CPUs.}
   \label{fig:scaling_res}
\end{figure}

Figure~\ref{fig:scaling_res} reports a modest-scale scaling test on a simulation with 17921 leaf octants, spreading across 4 increasingly refined mesh levels ($l_{max} = l_{ref} = 3$, $l_{add} = 0$). Each octant is 6 grid cells by 6 grid cells. The blocks on the finest level have 32 complex Fourier modes (i.e. 63 terms when evaluating Fourier expansion). We plot the time taken to evaluate one RK3 step against the number of processors used by the simulation in Figure~\ref{fig:scaling_res}.  The strong scaling is consistent with the corresponding fully inhomogeneous LGF method \cite{dorschner2020fast,yu2022multi}.  

As discussed previously \cite{liska2014parallel,yu2022multi}, the LGF approach to the Poisson inversion is extremely efficient given the complex and adaptive domain.  On a per-point basis, only purely FFT based algorithms are likely to be more efficient, but the required rectangular domain would waste many points for the flows we compute.  

For the specific case of incompressible flow solver with one Fourier diagonalizable direction, we can make a direct comparison with the Jacobi and block-Jacoci Preconditioned Conjugate Gradient methods (JPCG and bJPCG) employed in an unstructured-mesh solver \cite{borrell2011parallel}. The authors also reported solution times for a direct Schur-complement based decomposition method (DSD), but we refrain from comparisons as such a method requires precomputing the Cholesky factorization and would be prohibitive in an adaptive algorithm.   Our method is compared to JPCG and bJPCG in Table~\ref{tab:Computational_cost}, where the computational rate for the Poisson solution is reported\footnote{ The simulations by the bJPCG and JPCG are done using PowerPC 970MP 2.3GHz CPUs\cite{borrell2011parallel}. }. For our code, the test is performed in the context of the $Re=300$ and $Re=12,000$ cylinder flows\footnote{The $Re=300$ cylinder flow is simulated on the Bridges-2 supercomputer \cite{brown2021bridges} with AMD EPYC 7742 2.25GHz CPUs. The $Re=12,000$ cylinder flow is simulated on the Stampede-2 supercomputer \cite{stanzione2017stampede} using Intel Xeon Platinum 8380 2.3GHz CPUs. } to be discussed in Section~\ref{sec:validation}, which were computed using 256 and 2000 CPU cores, respectively.  Our algorithm is about an order of magnitude faster for $Re=300$ case, and 4 times faster for the $Re=12,000$ case.  The latter case was impacted by deteriorating parallel performance on the associated very large grid of about 400 million cells.

\begin{table}[]
    \centering
    \begin{tabular}{|l|l|}
    \hline
    Case & Computational Rate (cpu $\times \mu \textrm{s/pts}$) \\ \hline
    JPCG \cite{borrell2011parallel} $\xi = 0$     &   83.1   \\ \hline
    bJPCG \cite{borrell2011parallel} $\xi = 0$     &    79.8  \\ \hline
    Nek5000 \cite{fischer2007nek5000, hosseini2016direct} & 55.6 \\ \hline
    Present ($Re = 300$)     &  7.1    \\ \hline
    Present ($Re = 12,000$)     &   19.4   \\ \hline
    \end{tabular}
    \caption{Efficiency comparison between methods for the 2D screened Poisson problems (JPCG and bJPCG) and the 3D Poisson problem (Nek5000) in incompressible flow.  The JPCG and bJPCG values are based on those reported \cite{borrell2011parallel} for the $\xi=0$ parameter value in their screened Poisson problem, which represents the worst case. The Nek5000 value is based on the time to run one GMRES iteration \cite{hosseini2016direct} in their Poisson solver and the expected number of iterations for the GMRES algorithm to converge, estimated from the number of iterations for the JPCG method to converge \cite{borrell2011parallel}.}
    \label{tab:Computational_cost}
\end{table}

In addition, for the more general case of simulating incompressible external flows with one periodic direction, we can compare with the spectral element incompressible flow solver Nek5000 \cite{fischer2007nek5000} simulating the flow past a wing section \cite{hosseini2016direct}. In \cite{hosseini2016direct}, the runtime of one GMRES (generalized minimal residual method) iteration for their Poisson solver is reported\footnote{The GMRES iteration is computed on the Cray-XC40 computer Beskow at PDC (KTH) using Intel Xeon E5-2698v3 Haswell 2.3 GHz CPUs \cite{hosseini2016direct}.}. However, solving the Poisson equation requires many GMRES iterations. As indicated by \cite{borrell2011parallel}, the number of iterations for the JPCG method to solve the Poisson equation is 217. In addition, the conjugate gradient method (used by the JPCG method) has better convergence properties than GMRES \cite{trefethen2022numerical}. Thus, the number of iterations for the JPCG method to converge can serve as a lower-bound estimate of the number of iterations required for the GMRES algorithm to converge. As such, we compare the computational rate for Nek5000 to compute 217 GMRES iterations to the computational rate of our method, also in Table~\ref{tab:Computational_cost}. Our method is roughly 8 times faster than Nek5000 for the $Re=300$ case and roughly three times faster for the $Re = 12,000$ case. 

The computational efficiency reported here could potentially be further improved with enhancements to the parallelization strategy and other optimizations, but as it stands we believe our algorithm is competitive with (and potentially faster than) other state-of-the art incompressible flow solvers.

\section{Convergence Study}
\label{sec:convergence}

To study the convergence properties of our flow solver, we compute the flow past a cylinder with diameter $D$ and $Re=100$ using 16 complex Fourier coefficients (31 terms when evaluating Fourier series) and the following initial vorticity distributions:
\begin{equation}
    \omega_{n,z} = \exp({-n^2-|r|^2/D^2}).
\end{equation}
We obtain an initial velocity by solving the discrete 2D Poisson equation and the screened Poisson equations from the vorticity-streamfunction equation. We run the simulation for 1.024 $tU_\infty/D$ and used a uniform grid simulation with $\Delta x_0/D = 0.005$ as the reference solution. We consider cases where the base spatial resolutions $\Delta x_{base}$ satisfies $\Delta x_{base}/\Delta x_0 \in \{4, 8, 16, 32, 64\}$ and the finest level resolution $\Delta x_{fine}$ satisfies $\Delta x_{fine}/\Delta x_0 \ge 4$. In each mesh topology, $l_{max} = l_{ref} = N_l$, $l_{add} = 0$. On $l^{th}$ level, we refine a squared region centered at the origin with an edge length of ${3.84D}/{2^l}$. Mathematically, on $l^{th}$ level, we refine the region defined by the following set:
\begin{equation}
    \Omega_l^r = \left\{(x,y) : \left\lvert\frac{x}{D}\right\lvert < \frac{1.92}{2^l}, \left\lvert\frac{y}{D}\right\lvert < \frac{1.92}{2^l}\right\}.
\end{equation}

\begin{figure}
    \centering
    \includegraphics[width=0.8\textwidth]{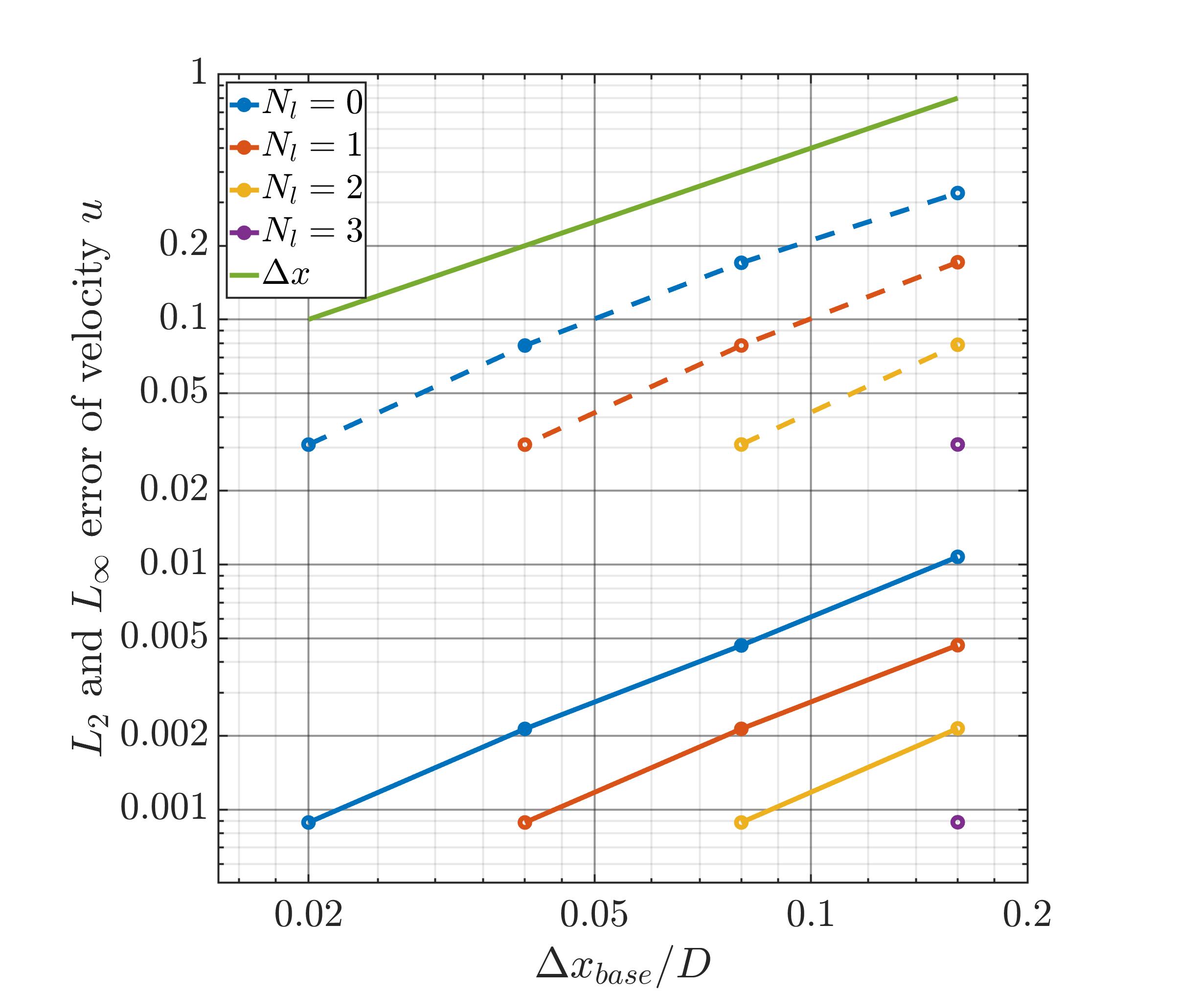}
    \caption{Error in the streamwise velocity compared to the base solution using different numbers of refinement levels ($0 \leq N_l \leq 3$).  The solid lines represent the $L_2$ error, and the dashed lines represent the $L_\infty$ error.}
    \label{fig:L2Err}
\end{figure}

The error is shown in Figure~\ref{fig:L2Err}.  We normalize $L_\infty$ error by the $L_\infty$ norm of the reference solution and $L_2$ error by the $L_\infty$ norm of the reference solution and the size of the domain. Both $L_\infty$ and $L_2$ error shows a first-order convergence, as expected for our 2nd-order finite-volume scheme with first-order immersed boundary method treatment \cite{tornberg2004numerical,mori2008convergence, taira2007immersed, colonius2008fast}.

\section{Validation}
\label{sec:validation}
\subsection{Cylinder flow at $Re = 300$}
We consider the flow over a circular cylinder at $Re = 300$, which has been extensively studied numerically and experimentally \cite{mittal1997inclusion, kravchenko1999b, norberg2003fluctuating}. At lower $Re$, the flow undergoes a series of bifurcations and at  300, the most prominent instability is termed mode-B and consists of vortex shedding modulated by three-dimensional streamwise vortex pairs. These form horseshoe-shaped vortices downstream that are stretched in the streamwise direction \cite{williamson1996vortex}. Furthermore, the number of horseshoe vortices decreases downstream due to a subharmonic instability \cite{mittal1997inclusion}. 

In our simulation, we use a spanwise period of $c = 12D$ where $D$ is the diameter of the cylinder, with 288 Fourier coefficients at the finest refinement level. Three levels of refinement ($l_{max} = l_{ref} = 3$, $l_{add} = 0$) are used, and thus the number of Fourier coefficients for computational cells on the coarsest level is 36. The base resolution is set to be $x_{base}/D = 0.08$. The mesh at each refinement level is increasingly refined with a factor of 2. Thus, the resolution on the finest level is $\Delta x_{3}/D = 0.01$. The adaptive mesh refinement algorithm locally refines and coarsens the computational domain with a refinement factor of $\alpha=0.25$ and a deletion factor of $\beta = 0.7$. The time step size is chosen as $\Delta tU_\infty/\Delta x_3 = 0.75$.

\begin{figure}
    \centering
    \includegraphics[width=0.8\textwidth]{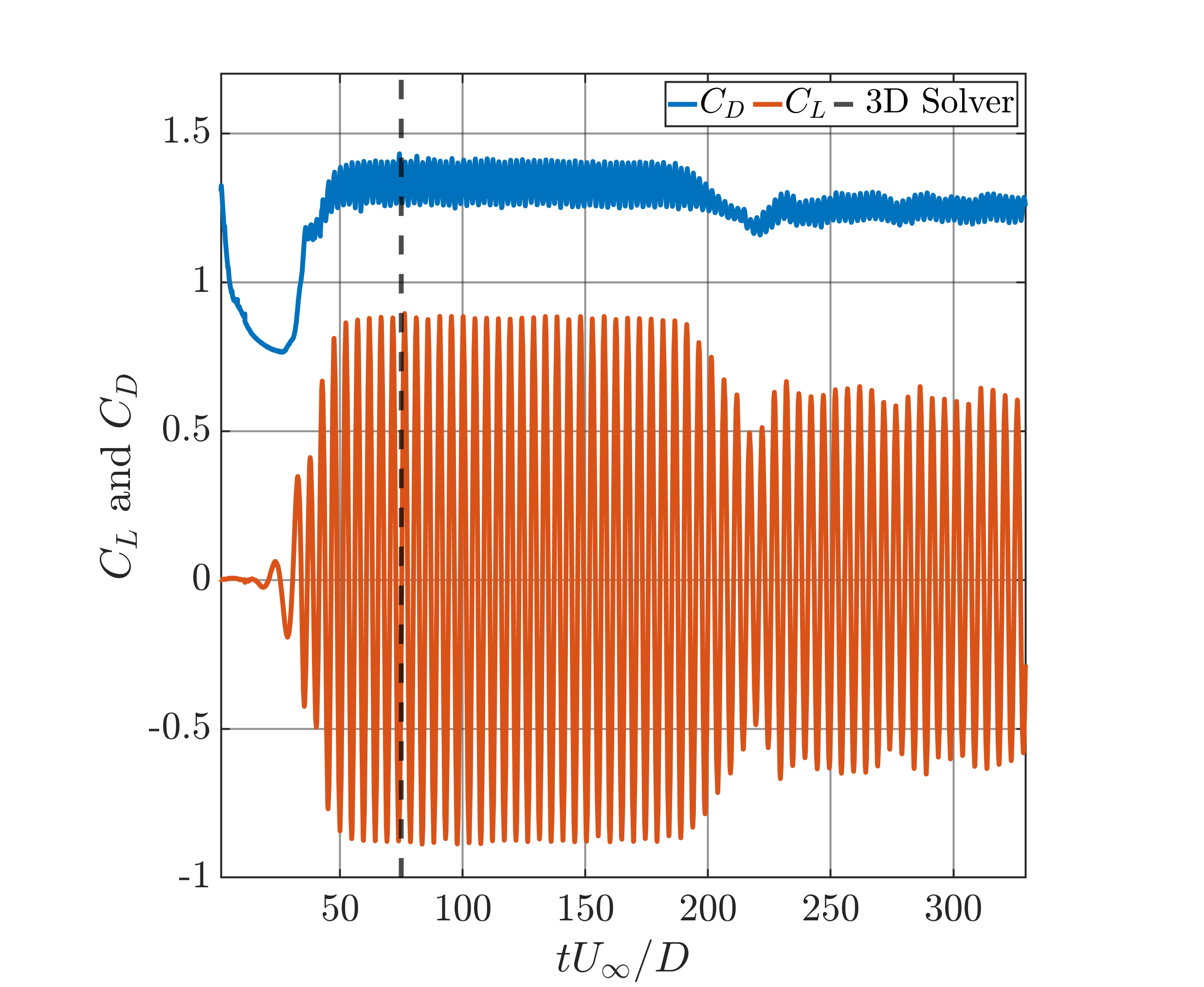}
    \caption{Lift and drag coefficient history of the simulation}
    \label{fig:LiftandDrag}
\end{figure}

To efficiently simulate this flow, we initialized the simulation by first computing the flow in 2D. After we reach a temporally periodic solution,  we initialize the 3D simulation using the 2D solution as the zeroth Fourier coefficient. At the beginning of the 3D simulation, a small (on the scale of $10^{-5}$) random vortical perturbation is introduced, with the expectation that the resulting flow becomes independent of the specific perturbation  \cite{mittal1997inclusion}. Integrating forward in time results in the lift $C_L$ and drag $C_D$ coefficients shown in Figure~\ref{fig:LiftandDrag}, where the dashed line denotes the initiation of the 3D simulation.

The three-dimensional instability is slow to develop, reaching a significant amplitude only by $tU_\infty/D = 175$ and saturating thereafter.  The flow is (approximately) stationary after $tU_\infty/D = 225$.  These dynamics are similar to what has been previously observed \cite{mittal1995effect, mittal1997inclusion}.

\begin{figure}
    \centering
    \includegraphics[width=\textwidth]{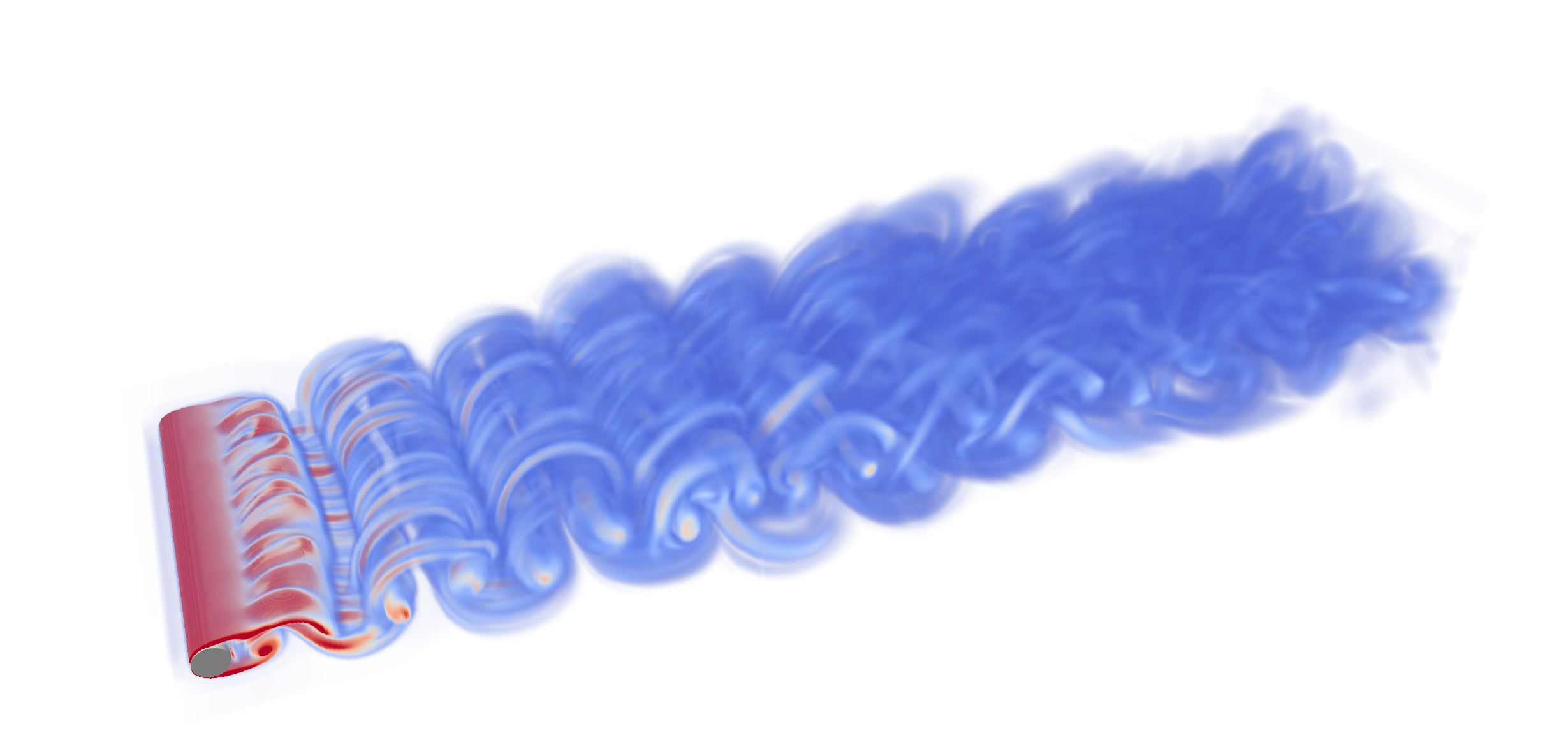}
    \caption{Vorticity magnitude at $tU_\infty/D = 367.5$. The non-dimensionalized vorticity, $\boldsymbol{\omega} D/U_\infty$, magnitude ranges from  $0$ (blue) to 5 (red).}
    \label{fig:Vort_mag}
\end{figure}

\begin{figure}
    \centering
    \includegraphics[width=\textwidth]{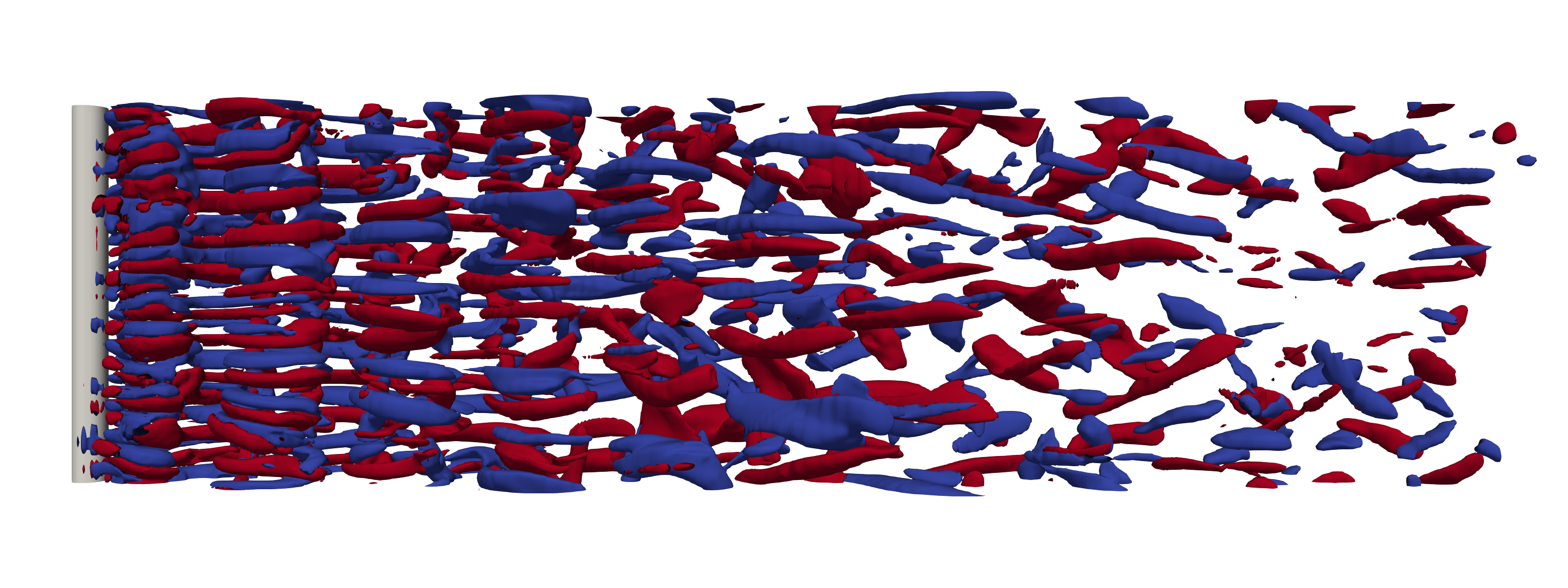}
    \caption{Streamwise vorticity contour plot at $tU_\infty/D = 367.5$ at $\omega_x D/U_\infty = 0.5$ (red) and $\omega_x D/U_\infty = -0.5$ (blue).}
    \label{fig:Vort_x}
\end{figure}

The vorticity magnitude and streamwise vorticity at $tU_\infty/D = 367.5$ are shown in figures~\ref{fig:Vort_mag} and \ref{fig:Vort_x}, respectively. We can clearly observe the three-dimension mode-B vortices forming in the wake near the cylinder and the elongated horseshoe vortices further downstream, with a continual decrease in the number of horseshoe vortices as the flow progresses. The formation of the horseshoe vortices and the reduction of the number of horseshoe vortices downstream indicates the subharmonic instability of the three-dimensional vortices generated by mode-B instability \cite{williamson1996vortex,mittal1997inclusion, kravchenko1999b}.


In addition to these qualitative observations, we computed the average drag coefficient ($\overline{C_D}$), root mean squared lift coefficient ($C_{L, rms}$), and Strouhal number ($St$) of this simulation. The mean is computed starting at $tU_\infty/D = 245.7$, which is more than 12 shedding cycles after the onset of three-dimensional instability. The statistics were accumulated for the next 14 shedding cycles. In Table~\ref{tab:Re300}, we report those statistics and compare the values to previous numerical and experimental studies. All quantities are in reasonable agreement with the reference values.

\begin{table}[]
\centering
\begin{tabular}{|l|l|l|l|}
\hline
Study & $\overline{C_D}$  & $C_{L, rms}$       & $St$       \\ \hline
Kravchenko et al. (num.) \cite{kravchenko1999b}   & 1.28 & 0.40 & 0.203 \\ \hline
Mittal and Balachandar (num.) \cite{mittal1997inclusion}   & 1.26 & 0.38 & 0.203 \\ \hline
Experimental \cite{wieselsberger1922further, norberg2003fluctuating}   & 1.22 & 0.45 & 0.203 \\ \hline
Present   & 1.25 & 0.44 & 0.203 \\ \hline
\end{tabular}
\caption{Comparison of lift and drag statistics with previous studies.}
\label{tab:Re300}
\end{table}

\subsection{Cylinder flow at $Re = 12,000$}

We now consider the fully turbulent cylinder flow at $Re = 12,000$. We carried out the simulation using 128 Fourier coefficients and spanwise periodic length $c = 3D$. In this simulation, we used a base resolution $\Delta x_{base}/D = 0.04$ with four levels of adaptive refinement, yielding the finest resolution of $\Delta x_4/D = 0.0025$. The refinement factor is $\alpha=0.25$, and the deletion factor is $\beta = 0.7$. At $Re = 3,900$, the average Kolmogorov scale ($\bar{\eta}$) in the near wake ($x/D < 5$) is $\bar{\eta}/D = 0.02$ \cite{lehmkuhl2013low}. According to the $3/4$ scaling of the Kolmogorov scale, we estimate that $\bar{\eta}/D = 0.0086$ at $Re = 12,000$. The ratio of the Kolmogorov scale to the second finest level is $\Delta x_3/\bar{\eta} = 0.58$. Thus, the turbulence is expected to be fully resolved.  The finest level is required to resolve the thin laminar boundary layers on the cylinder prior to separation and the shear layer attached to the cylinder. 

To efficiently reach the fully developed turbulent flow, we adopted a step-wise strategy. Specifically, we first initialize a 2D simulation of flow past a cylinder at $Re = 5,000$. After vortex shedding initiates, we use that flow profile as the zeroth Fourier coefficient to initialize a fully 3D simulation with 64 spanwise Fourier coefficients. We perturb this 3D simulation with a small random vortical perturbation (on the level $10^{-5}$) to trigger the spanwise instability. After the flow becomes fully turbulent (at $tU_\infty/D \approx 25$), we increase the Reynolds number to 12,000 and increase the number of spanwise Fourier coefficients to 128 to continue our simulation. The simulation is then continued for more than $125tU_\infty/D$. Throughout the simulation, the time step size is chosen to be $\Delta t U_\infty/\Delta x = 0.5$. The value is chosen to satisfy the CFL criterion \cite{courant1928partiellen}. 

\begin{figure}
    \centering
    \includegraphics[width=\textwidth]{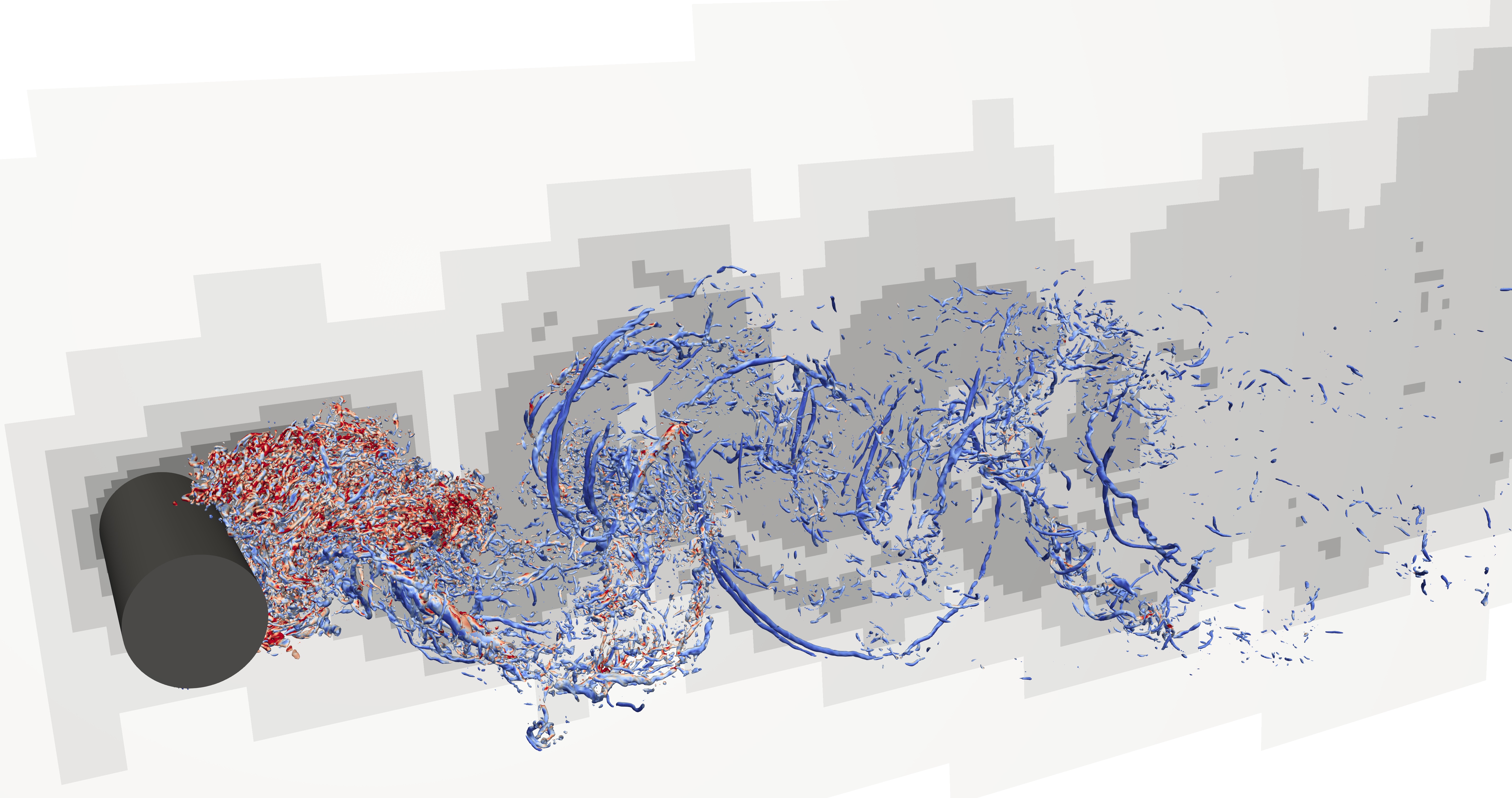}
    \caption{Q-criterion isosurface at $Q = 100 U_\infty/D$. The isosurface is colored by the vorticity magnitude from $||\boldsymbol{\omega}||D/U_\infty = 20$ (blue) to $50$ (red). The corresponding $x-y$ computational mesh is shown in the background. The mesh gets increasingly fine as the gray darkens. The computational domain is truncated in this figure. The full computational domain is adaptive and extends to 23D downstream.}
    \label{fig:Q_cr}
\end{figure}

\begin{figure}
    \centering
    \includegraphics[width = 0.5\textwidth]{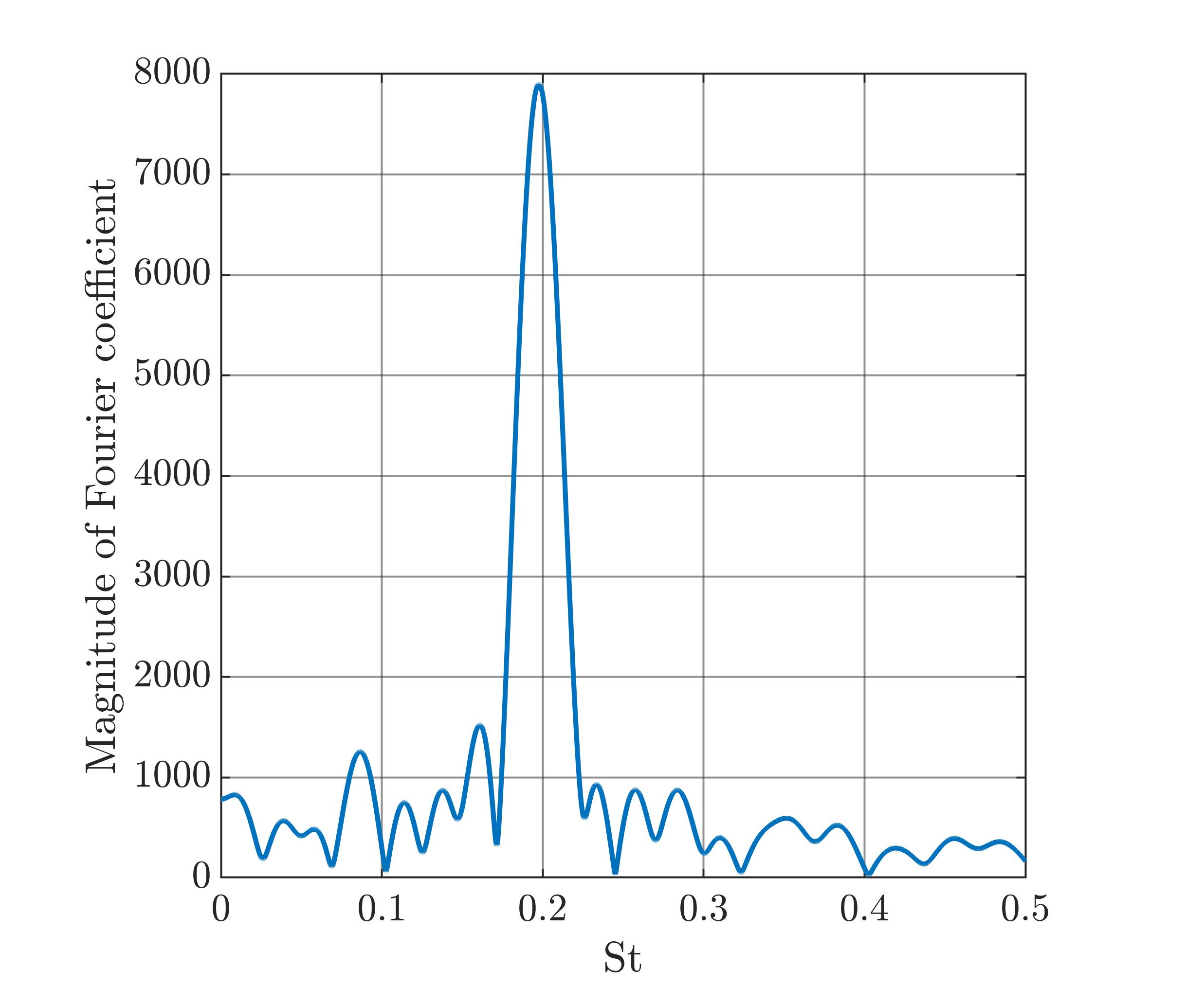}
    \caption{Fourier spectrum of lift coefficient.}
    \label{fig:FFT_cl}
\end{figure}

Figure~\ref{fig:Q_cr} shows the isosurface of the Q-criterion at $tU_\infty/D = 145.75 $. We can clearly observe the rib structures in the wake that are also present in the wake of the flow past a flat plate \cite{rai2013flow}. We also show the $x-y$ mesh topology in the same figure. The computational grid tracks the vortical region of the flow and adaptively refines in the high vorticity region. We computed statistics using the data within the time period $tU_\infty/D \in [105, 145]$, corresponding to approximately 8 vortex-shedding cycles. When estimating a Strouhal number, we conduct a Fourier transform of the lift coefficient data in time. The resulting Fourier spectrum is shown in Figure~\ref{fig:FFT_cl}. The peak of the Fourier spectrum centers around $St = 0.198$.

\begin{figure}
    \centering
    \includegraphics[width=0.7\textwidth]{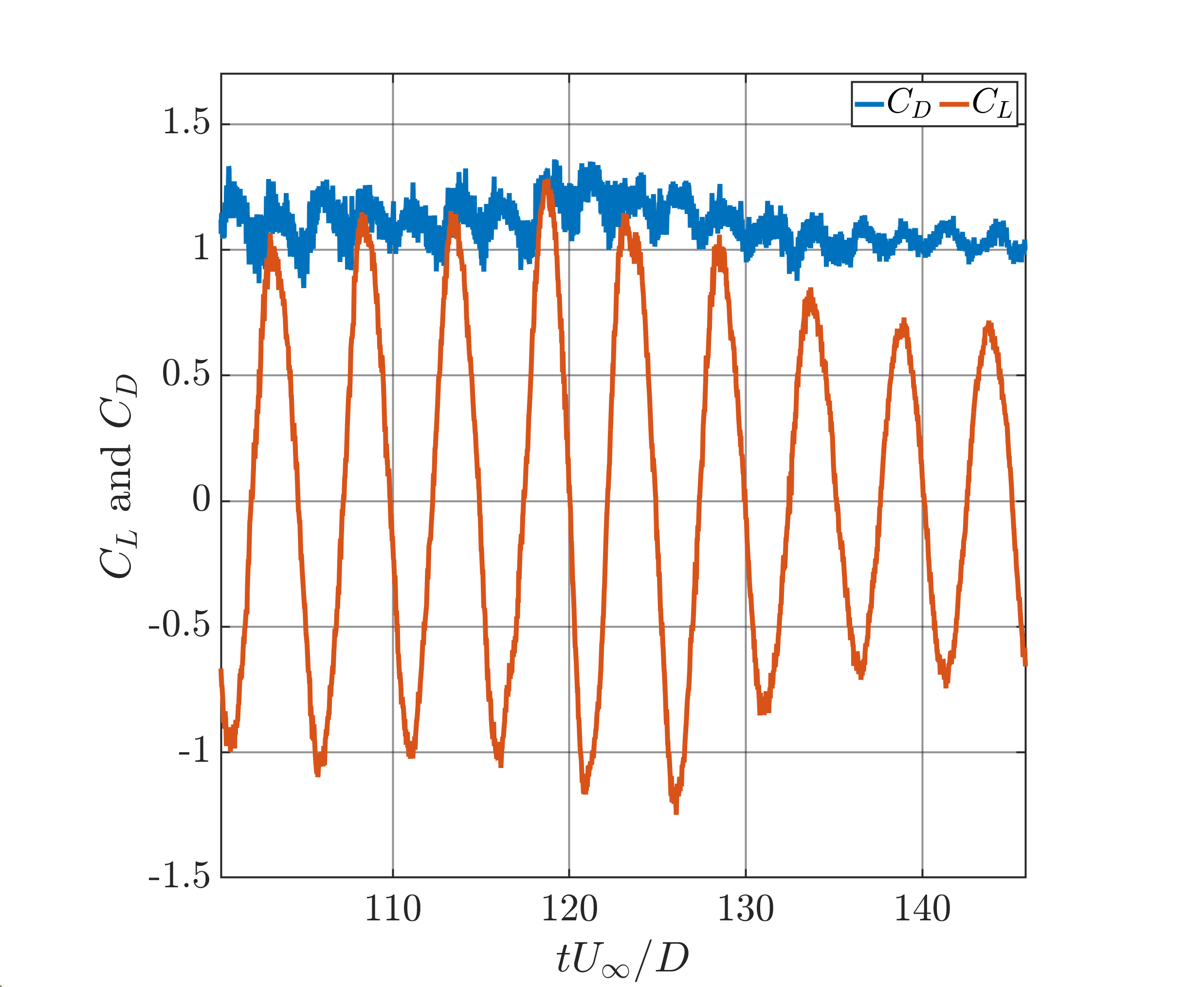}
    \caption{Drag coefficient and lift coefficient time evolution during $tU_\infty/D  \in [100, 145]$.}
    \label{fig:Re12000_clcd}
\end{figure}

The drag coefficient and lift coefficient time evolution data for $tU_\infty/D \ \in [100, 145]$ are shown in Figure~\ref{fig:Re12000_clcd}. The mean drag and Strouhal number are compared to experimental data in Table~\ref{tab:comp12000}. The Strouhal number and $C_{L, rms}$ are computed from an empirical formula obtained from experimental data \cite{norberg2003fluctuating} when $Re \in [1600, 150000]$ for $St$ and $Re \in [5400, 220000]$ for $C_{L, rms}$:
\begin{equation}
    St = 0.1853 + 0.0261 \times \exp (-0.9 \times x^{2.3}), \qquad x = \log (Re/1600),
\end{equation}
\begin{equation}
    C_{L, rms} = 0.52 - 0.06 \times x^{-2.6}, \qquad x = \log (Re/1600).
\end{equation}
\begin{table}[]
\centering
\begin{tabular}{|l|l|l|l|}
\hline
Study & $\overline{C_D}$  & $C_{L, rms}$       & $St$       \\ \hline
Norberg (exp.) \cite{norberg2003fluctuating}   & - & 0.435 & 0.199 \\ \hline
Wieselsberger (exp.) \cite{wieselsberger1922further}   & 1.15 & - & - \\ \hline
Present   & 1.12 & 0.67 & 0.198 \\ \hline
\end{tabular}
\caption{Drag coefficient, lift coefficient, and Strouhal number comparison between present numerical method and experimental data for the flow past a cylinder at $Re=12,000$.}
\label{tab:comp12000}
\end{table}
The drag coefficient and Strouhal number agree with the reference values within uncertainty associated with the empirical formula (see Figures 1 and 2 in \cite{norberg2003fluctuating}).  On the other hand, $C_{L,rms}$ is about 50\% higher than the formula.  As can be observed in Figure~\ref{fig:Re12000_clcd}, there is a more significant variation in the lift from cycle to cycle, and the discrepancy is likely attributable to the statistic not having converged.

\section{Summary}
\label{sec:conclusion}

We have extended the IB-LGF-AMR approach previously developed for unbounded three-dimensional flows to solve external flows around cylinders of arbitrary cross-sections. To this end, we hybridized a mimetic, staggered-mesh finite-volume discretization for the inhomogeneous directions with a Fourier spectral method for the homogeneous spanwise direction, using a pseudo-spectral approach for the nonlinear term.   Some innovations include the implementation of AMR on the Fourier coefficients, efficient direct LU solution of the system of equations for the IB forcing, and new algorithms for efficiently tabulating the associated LGFs for the screened Poisson operator and viscous integrating factor to high precision.  The validated algorithm retains the flexibility, efficiency, scalability, and accuracy of the IB-LGF-AMR approach, as demonstrated by computing the fully turbulent flow by past a circular cylinder at $Re=12,000$.

\section*{Acknowledgements}
This work was supported in part by The Boeing Company (CT-BA-GTA-1). We gratefully acknowledge computational allocations on the Stampede2 and Bridges-2 supercomputers awarded by the XSEDE program with allocation number CTS120005 and the ACCESS program with allocation number PHY230039. We would also like to acknowledge the help from Ke Yu and Benedikt Dorschner during the development of this algorithm.

\appendix
\section{IB terms for Fourier Coefficients}
\label{ap:NS_ib_Fourier}

In this part, we provide a few details regarding the Fourier transform of Eq.\ref{eq:NS_ib} leading to Eq.\ref{eq:FT_NS_ib}. The only terms that are nontrivial are the Fourier transforms of the IB forcing term and the no-slip boundary condition.  

We first show the derivations of the IB forcing term. Consider the immersed boundary defined, $\Gamma(t)$. The geometry of the immersed boundary is assumed to be the extrusion of a two-dimensional boundary. Thus, we can denote that said two-dimensional boundary as $\Gamma_{2D}(t)$. With this, we can write:
\begin{equation}
    \Gamma(t) = \Gamma_{2D}(t) \times [-c/2, c/2].
\end{equation}
Consequently, we can find some $\boldsymbol{\xi}_{2D}$ that parameterizes $\Gamma_{2D}(t)$ for all $t$. Correspondingly, the parameterization at time $t$ is defined as $\boldsymbol{X}_{2D}(\boldsymbol{\xi}_{2D}, t)$. Thus, following the notation of Eq.~\ref{eq:FT_NS_ib}, we can write
\begin{equation}
    \boldsymbol{X}(\boldsymbol{\xi},t) = (\boldsymbol{X}_{2D}(\boldsymbol{\xi}_{2D}, t), Z), \quad \boldsymbol{\xi} = (\boldsymbol{\xi}_{2D}, Z).
\end{equation}
Then with $\boldsymbol{x} = (\boldsymbol{x}_{2D}, z)$, we can write the IB forcing term at $\boldsymbol{x}$ as:
\begin{align*}
    &\int_{\Gamma(t)} \boldsymbol{f}_\Gamma(\boldsymbol{\xi}, t) \delta(\boldsymbol{X}(\boldsymbol{\xi}, t) - \boldsymbol{x}) d\boldsymbol{\xi} \\
    &= \int_{-c/2}^{c/2}\int_{\Gamma_{2D}(t)} \boldsymbol{f}_\Gamma(\boldsymbol{\xi}_{2D}, Z, t) \delta_{2D}(\boldsymbol{X}_{2D}(\boldsymbol{\xi}, t) - \boldsymbol{x}_{2D}) \delta_{1D}(z-Z)d\boldsymbol{\xi}_{2D}dZ
\end{align*}
where $\delta_{1D}$ and $\delta_{2D}$ denote the Delta functions in 1D spaces and 2D spaces, respectively. With the above notation, we write:
\begin{equation}
    \begin{aligned}
    &\mathcal{F}_k \left[ \int_{\Gamma(t)} \boldsymbol{f}_\Gamma(\boldsymbol{\xi}, t) \delta(\boldsymbol{X}(\boldsymbol{\xi}, t) - \boldsymbol{x}) d\boldsymbol{\xi} \right] \\
    = &\frac{1}{c}\int_{-c/2}^{c/2} e^{-iz\frac{2\pi k}{c}} \int_{\Gamma(t)} \boldsymbol{f}_\Gamma(\boldsymbol{\xi}, t) \delta(\boldsymbol{X}(\boldsymbol{\xi}, t) - \boldsymbol{x}) d\boldsymbol{\xi} dz \\
    = &\frac{1}{c}\int_{\Gamma(t)} \int_{-c/2}^{c/2}e^{-iz\frac{2\pi k}{c}} \boldsymbol{f}_\Gamma(\boldsymbol{\xi}, t) \delta(\boldsymbol{X}(\boldsymbol{\xi}, t) - \boldsymbol{x}) dz d\boldsymbol{\xi} \\
    = &\frac{1}{c}\int_{\Gamma_{2D}(t)} \int_{-c/2}^{c/2} \int_{-c/2}^{c/2} e^{-iz\frac{2\pi k}{c}} \boldsymbol{f}_\Gamma(\boldsymbol{\xi}_{2D}, Z, t) \delta_{2D}(\boldsymbol{X}_{2D}(\boldsymbol{\xi}_{2D}, t) - \boldsymbol{x}_{2D}) \delta_{1D}(z - Z) dZ dz d\boldsymbol{\xi}_{2D} \\
    = &\frac{1}{c}\int_{\Gamma_{2D}(t)} \int_{-c/2}^{c/2} e^{-iz\frac{2\pi k}{c}} \boldsymbol{f}_\Gamma(\boldsymbol{\xi}_{2D}, z, t) \delta_{2D}(\boldsymbol{X}_{2D}(\boldsymbol{\xi}_{2D}, t) - \boldsymbol{x}_{2D}) dz d\boldsymbol{\xi}_{2D} \\
    = &\int_{\Gamma_{2D}(t)} \tilde{\boldsymbol{f}}_{\Gamma, k} (\boldsymbol{\xi}_{2D}, t)\delta_{2D}(\boldsymbol{X}_{2D}(\boldsymbol{\xi}_{2D}, t) - \boldsymbol{x}_{2D}) d\boldsymbol{\xi}_{2D},
    \end{aligned} 
\end{equation}
where $\tilde{f}_{\Gamma,k}$ is the $k^{th}$ Fourier coefficient of the IB forcing $f_\Gamma$. 

For the boundary condition, we apply $\mathcal{F}_k$ directly. The LHS, by definition, is:
\begin{equation}
    \begin{aligned}
    &\mathcal{F}_k \left[ \boldsymbol{u}_\Gamma(\boldsymbol{\xi},t) \right]\\
    = &\frac{1}{c}\int_{-c/2}^{c/2} e^{-iZ\frac{2\pi k}{c}} \boldsymbol{u}_\Gamma(\boldsymbol{\xi},t) dZ \\
    = &\frac{1}{c}\int_{-c/2}^{c/2} e^{-iZ\frac{2\pi k}{c}} \boldsymbol{u}_\Gamma(\boldsymbol{\xi}_{2D}, Z, t) dZ \\
    = &\widetilde{\boldsymbol{u}}_{\Gamma, k}(\boldsymbol{\xi}_{2D}, t).
    \end{aligned}
\end{equation}
The RHS is:
\begin{equation}
    \begin{aligned}
    &\mathcal{F}_k \left[ \int_{\mathbb{R}^3}\boldsymbol{u}(\boldsymbol{x},t)\delta(\boldsymbol{x} - \boldsymbol{X}(\boldsymbol{\xi}, t)) d\boldsymbol{x} \right] \\
    = &\frac{1}{c}\int_{-c/2}^{c/2} e^{-iZ\frac{2\pi k}{c}} \int_{\mathbb{R}^3}\boldsymbol{u}(\boldsymbol{x},t)\delta(\boldsymbol{x} - \boldsymbol{X}(\boldsymbol{\xi}, t)) d\boldsymbol{x} dZ \\
    = &\frac{1}{c}\int_{-c/2}^{c/2} e^{-iZ\frac{2\pi k}{c}} \int_{\mathbb{R}} \int_{\mathbb{R}^2}\boldsymbol{u}(\boldsymbol{x}_{2D}, z,t)\delta_{2D}(\boldsymbol{x}_{2D} - \boldsymbol{X}_{2D}(\boldsymbol{\xi}_{2D}, t))\delta_{1D}(z-Z) d\boldsymbol{x}_{2D}dz dZ \\
    = &\frac{1}{c}\int_{-c/2}^{c/2} e^{-iZ\frac{2\pi k}{c}} \int_{\mathbb{R}^2}\int_{\mathbb{R}} \boldsymbol{u}(\boldsymbol{x}_{2D}, z,t)\delta_{2D}(\boldsymbol{x}_{2D} - \boldsymbol{X}_{2D}(\boldsymbol{\xi}_{2D}, t))\delta_{1D}(z-Z) dz d\boldsymbol{x}_{2D} dZ \\
    = &\frac{1}{c}\int_{-c/2}^{c/2} e^{-iZ\frac{2\pi k}{c}} \int_{\mathbb{R}^2} \boldsymbol{u}(\boldsymbol{x}_{2D}, Z,t)\delta_{2D}(\boldsymbol{x}_{2D} - \boldsymbol{X}_{2D}(\boldsymbol{\xi}_{2D}, t)) d\boldsymbol{x}_{2D} dZ \\
    = &\int_{\mathbb{R}^2} \left[ \frac{1}{c}\int_{-c/2}^{c/2} e^{-iZ\frac{2\pi k}{c}}  \boldsymbol{u}(\boldsymbol{x}_{2D}, Z,t) dZ\right]\delta_{2D}(\boldsymbol{x}_{2D} - \boldsymbol{X}_{2D}(\boldsymbol{\xi}_{2D}, t)) d\boldsymbol{x}_{2D} \\
    = &\int_{\mathbb{R}^2}{\tilde{\boldsymbol{u}}}_k(\boldsymbol{x}_{2D},t)\delta_{2D}(\boldsymbol{x}_{2D} - \boldsymbol{X}_{2D}(\boldsymbol{\xi}_{2D}, t))d\boldsymbol{x}_{2D} .
    \end{aligned} 
\end{equation}
Then, the boundary condition becomes:
\begin{equation}
\begin{aligned}
    \widetilde{\boldsymbol{u}}_{\Gamma, k}(\boldsymbol{\xi}_{2D}, t) &= \int_{\mathbb{R}^2}{\tilde{\boldsymbol{u}}}_k(\boldsymbol{x}_{2D},t)\delta_{2D}(\boldsymbol{x}_{2D} - \boldsymbol{X}_{2D}(\boldsymbol{\xi}_{2D}, t))d\boldsymbol{x}_{2D} .
\end{aligned}
\end{equation}

\section{Convergence Rate of Trapoziodal Appximation}
\label{ap:TrapConv}
We start with a theorem from Trefethen and Weideman \cite{trefethen2014exponentially}.
\begin{theorem*}
Let
\begin{equation}
    I = \int_{-\pi}^{\pi}v(\theta)d\theta
\end{equation}
For any positive integer $N$, define the trapezoidal rule approximation:
\begin{equation}
    I_N = \frac{2\pi}{N}\sum\limits_{k=1}^N v(\theta_k)
\end{equation}
where $\theta_k = 2\pi k/N - \pi$. Suppose $v$ is $2\pi$ periodic and analytic and satisfies $|v(\theta)| < M$ in the strip $-\gamma < \Im(\theta) < \gamma$ for some $\gamma > 0$. Then for any $N \geq 1$,
    \begin{equation}
        |I_N - I| \leq \frac{4\pi M}{e^{\gamma N} - 1},
        \label{eq:converate_Tref}
    \end{equation}
    and the constant $4\pi$ is as small as possible.
\end{theorem*}
In our specific case, 
\begin{equation}
\begin{aligned}
    &v(\theta) = (1 - \frac{e^{i\theta m}}{K^{|n|}}) \frac{1}{K - 1/K},\\
    &a = 4 + \left(\frac{2\pi k\Delta x}{c} \right)^2 - 2\cos(\theta), \\
    &K = \frac{a + \sqrt{a^2 - 4}}{2}.
\end{aligned}
\end{equation}
So we only need to pick any finite $\gamma$ such that 
\begin{equation}
    K - 1/K \neq 0 \textrm{ and } a^2 - 4 \notin \mathbb{R}_{\leq 0}\quad \forall \theta \in \mathbb{C}: |\Im(\theta)| < \gamma.
\end{equation}
We have 
\begin{equation}
    K - 1/K = \sqrt{a^2 - 4},
\end{equation}
so 
\begin{equation}
    K - 1/K = 0 \iff a^2 = 4 \iff a = \pm 2.
\end{equation}
Directly plugging in, we obtain:
\begin{equation}
    \cos(\theta) = 2 \pm 1 + \frac{1}{2}\left(\frac{2\pi k\Delta x}{c} \right)^2.
\end{equation}
Let
\begin{equation}
    \phi^+_m = 3 + \frac{1}{2}\left(\frac{2\pi k\Delta x}{c} \right)^2, \quad \phi^-_m = 1 + \frac{1}{2}\left(\frac{2\pi k\Delta x}{c} \right)^2.
\end{equation}
We can find this is true only when 
\begin{equation}
    \exp({\Im(\theta)}) = \phi^+_m \pm \sqrt{(\phi^+_m)^2 - 1} \textrm{ or } \exp{(\Im(\theta))} = \phi^-_m \pm \sqrt{(\phi^-_m)^2 - 1}
\end{equation}
Since logarithm is a monotonically increasing function and that $|\log(\phi^\pm_m + \sqrt{(\phi^\pm_m)^2 - 1})| = |\log(\phi^\pm_m - \sqrt{(\phi^\pm_m)^2 - 1})|$, to ensure analyticity within the strip, we need 
\begin{equation}
    |\Im(\theta)| < \log(\phi^-_m + \sqrt{(\phi^-_m)^2 - 1}).
\end{equation}
We also verify that when the above condition is satisfied, 
\begin{equation}
    |\Re(\cos(\theta))| \leq  |\cos(\theta)| < \phi^-_m = 1 + \frac{1}{2}\left(\frac{2\pi k \Delta x}{c}\right)^2.
\end{equation}
Thus, we can lower bound the real part of $a$ as $\Re(a) > 2$. For $a^2 - 4 \in \mathbb{R}_{\leq 0}$ to be true, we can only have two possibilities: either $a^2 \in [0,4]$ or $a^2 < 0$. Both cases are not possible since $\Re(a) > 2$. 

Now, we have proved that $v(\theta)$ is analytic if the assumption below is satisfied:
\begin{equation}
    |\Im(\theta)| < \gamma := \log(\phi^-_m + \sqrt{(\phi^-_m)^2 - 1}).
\end{equation}
Thus, if we pick any $\gamma_c < \gamma$, we have
\begin{equation}
    |v(\theta)| \leq M = \sup\limits_{|\Im(\theta)| < \gamma_c} \left| (1 - \frac{e^{i\theta m}}{K^{|n|}})\frac{1}{K - 1/K} \right|.
\end{equation}
Further, we can write
\begin{equation}
    M \leq \sup\limits_{|\Im(\theta)| < \gamma_c} \left| (1 - \frac{1}{K^{|n|}})\frac{1}{K - 1/K} \right| + \sup\limits_{|\Im(\theta)| < \gamma_c} \left| \frac{e^{i\theta m} - 1}{K^{|n|}}\frac{1}{K - 1/K} \right|.
\end{equation}
The first term is the $M$ for $m = 0$, the second term is bounded by $C(\exp(m\gamma_c) - 1)$ for some $C$ independent of $m$. Combined with Eq.~\ref{eq:converate_Tref}, increasing $m$ does not affect the convergence rate in the number of quadrature points in the asymptotic sense but only change the constant $M$ in the convergence rate.

\section{Compatibility Condition on $L_0^{-1}$ in a Multi-resolution Mesh}
\label{ap:LGF_compat}
Consider two different uniform Cartesian meshes with different resolutions $\Delta x_1$ and $\Delta x_2$ and are governed by the relationship $\Delta x_1 = N \Delta x_2$, where $N$ a positive integer. Denote the $B_0$ kernel for the two grid as $B_0^{1}$ and $B_0^2$. The compatibility condition we impose is:
\begin{equation}
    \lim\limits_{|\boldsymbol{n}| \rightarrow \infty} (B_0^1(\boldsymbol{n}) - B_0^2(\boldsymbol{n}N)) = 0.
    \label{eq:compat_2DLGF}
\end{equation}
Since $\boldsymbol{n}\Delta x_1 = \boldsymbol{n}N\Delta x_2$, this condition means that, if $B_0^1$ and $B_0^2$ are two different discrete solutions of a discretized 2D Poisson equation induced by the Dirac delta function, both solutions should match at infinity in the physical space. 

Using the asymptotic expansion of LGF \cite{martinsson2002asymptotic}, we can write:
\begin{equation}
    B_0(\boldsymbol{n}) = \frac{1}{2\pi}\log(|\boldsymbol{n}|) + C + O(1/|\boldsymbol{n}|^2).
\end{equation}
Thus, by plugging in this expansion to Eq.~\ref{eq:compat_2DLGF}, we get:
\begin{equation}
    \begin{aligned}
        0 &= \lim\limits_{|\boldsymbol{n}| \rightarrow \infty} (B_0^1(\boldsymbol{n}) - B_0^2(\boldsymbol{n}N)) \\
          &= \lim\limits_{|\boldsymbol{n}| \rightarrow \infty} (\frac{1}{2\pi}\log(|\boldsymbol{n}|) + C_1 + O(1/|\boldsymbol{n}|^2) - \frac{1}{2\pi}\log(N|\boldsymbol{n}|) - C_2 - O(1/|N\boldsymbol{n}|^2)) \\
          &= \lim\limits_{|\boldsymbol{n}| \rightarrow \infty} (-\frac{1}{2\pi}\log(N) + C_1 -C_2 + O(1/|\boldsymbol{n}|^2)) \\
          &= -\frac{1}{2\pi}\log(N) + C_1 -C_2.
    \end{aligned}
\end{equation}
Thus, we obtain 
\begin{equation}
    C_1 - C_2= \frac{1}{2\pi}\log(N).
\end{equation}
That is:
\begin{equation}
    B_0^1(\boldsymbol{n}) = B_0^2(\boldsymbol{n}) + \frac{1}{2\pi}\log(N).
\end{equation}
In the context of our algorithm, we have $N = 2^l$ for some non-negative integer $l$. As a result, we have:
\begin{equation}
    B_0^1(\boldsymbol{n}) = B_0^2(\boldsymbol{n}) + \frac{l}{2\pi}\log(2).
\end{equation}
\section{$S^i_{k,n}$ is Hermitian when $P^i_n = P^{i-1}_n$}
\label{ap:S_hermitian}
Here we show that $(S_{k,n}^i)^* = S_{k,n}^i$ under the assumption of $P^i_n = P^{i-1}_n = P_n$. In this case, we have $S^i_{k,n} = P_{n}E^i_{k}(I - G_kL_k^{-1}D_k)P_n^T$.
Since $E^i_{k,n}$ commutes with all the discrete operators and $L_k^{-1}$, we have:
\begin{equation}
\begin{aligned}
    (S_{k,n}^i)^* &= \left[P_{n}E^i_{k}(I - G_kL_k^{-1}D_k)P_n^T\right]^* \\
    &= (P_{n}^T)^*(E^i_{k})^*(I - G_kL_k^{-1}D_k)^*P_n^*. \\
    \end{aligned}
    \end{equation}
Since $P_n$ is purely real, $P_n^* = P_n^T$. In addition, we know that $(E^i_{k})^* = E^i_{k}$, $(L_k^{-1})^* = L_k^{-1}$. As a result, we can write the equation above as:
\begin{equation}
     (P_{n}^T)^*(E^i_{k})^*(I - G_kL_k^{-1}D_k)^*P_n^* = P_{n}E^i_{k}(I - D_k^*L_k^{-1}G_k^*)P_n^T.
\end{equation}
Now we recall the mimetic property $D_k^* = -G_k$, and we obtain that:
\begin{equation}
    P_{n}E^i_{k}(I - D_k^*L_k^{-1}G_k^*)P_n^T = P_{n}E^i_{k}(I - G_kL_k^{-1}D_k)P_n^T.
\end{equation}
Combing all the steps together, we have shown that:
\begin{equation}
\begin{aligned}
    (S_{k,n}^i)^* &= \left[P_{n}E^i_{k}(I - G_kL_k^{-1}D_k)P_n^T\right]^* \\
    &= (P_{n}^T)^*(E^i_{k})^*(I - G_kL_k^{-1}D_k)^*P_n^* \\
    &= P_{n}E^i_{k}(I - D_k^*L_k^{-1}G_k^*)P_n^T \\
    &= P_{n}E^i_{k}(I - G_kL_k^{-1}D_k)P_n^T \\
    &= S^i_{k,n}.
\end{aligned}
\end{equation}
Thus, we conclude that $S^i_{k,n}$ is Hermitian.

 \bibliographystyle{elsarticle-num} 
 \bibliography{elsarticle-template-num}

\end{document}